\documentclass[12pt]{article}

\newcommand{\Om}{\Omega}

\newcommand{\al}{\alpha}
\newcommand{\ep}{\epsilon}

\newcommand{\NS}{\mbox{NS}}
\newcommand{\tNS}{\widetilde{\mbox{NS}}}
\newcommand{\R}{\mbox{R}}
\newcommand{\tR}{\widetilde{\mbox{R}}}
\newcommand{\sNS}{\msc{NS}}
\newcommand{\stNS}{\widetilde{\msc{NS}}}
\newcommand{\sR}{\msc{R}}
\newcommand{\stR}{\widetilde{\msc{R}}}

\newcommand{\deebar}{\bar{\partial}}

\newcommand{\lb}{\lbrack}
\newcommand{\rb}{\rbrack}

\newcommand{\msc}[1]{\mbox{\scriptsize #1}}
\newcommand{\dsp}{\displaystyle}
\newcommand{\scs}[1]{{\scriptstyle #1}}

\newcommand{\bc}{\mbox{{\bf C}}}
\newcommand{\br}{\mbox{{\bf R}}}
\newcommand{\bz}{\mbox{{\bf Z}}}

\newcommand{\bsz}{\msc{{\bf Z}}}
\newcommand{\bsr}{\msc{{\bf R}}}

\newcommand{\cG}{{\cal G}}
\newcommand{\cJ}{{\cal J}}

\newcommand{\cN}{{\cal N}}
\newcommand{\cM}{{\cal M}}
\newcommand{\cF}{{\cal F}}

\newcommand{\cD}{{\cal D}}

\newcommand{\cH}{{\cal H}}
\newcommand{\cU}{{\cal U}}
\newcommand{\cZ}{{\cal Z}}
\newcommand{\cI}{{\cal I}}
\newcommand{\cK}{{\cal K}}

\newcommand{\tL}{\tilde{L}}
\newcommand{\tJ}{\tilde{J}}
\newcommand{\tI}{\tilde{I}}

\newcommand{\tF}{\tilde{F}}
\newcommand{\tN}{\tilde{N}}

\newcommand{\tj}{\tilde{j}}
\newcommand{\tm}{\tilde{m}}
\newcommand{\tl}{\tilde{l}}

\newcommand{\hC}{\hat{\cal C}}
\newcommand{\hD}{\hat{\cal D}}
\newcommand{\hE}{\hat{\cal E}}

\newcommand{\ket}[1]{{|#1\rangle}}

\newcommand{\Th}[2]{\Theta_{#1,#2}}
\renewcommand{\th}{{\theta}}
\newcommand{\tTh}[2]{\tilde{\Theta}_{#1,#2}}
\newcommand{\ch}[2]{\mbox{ch}^{#1}_{#2}}

\newcommand{\Ch}[2]{\mbox{Ch}^{#1}_{#2}}

\newcommand{\chm}[1]{\mbox{ch}^{#1}_{\msc{\bf M}}}
\newcommand{\chg}[1]{\mbox{ch}^{#1}_{\msc{\bf G}}}

\newcommand{\chic}{{\chi_{\msc{\bf con}}}}
\newcommand{\chid}{{\chi_{\msc{\bf dis}}}}

\newcommand{\Zc}{{Z_{\msc{\bf con}}}}
\newcommand{\Zd}{{Z_{\msc{\bf dis}}}}

\newcommand{\Chm}[1]{\mbox{Ch}^{#1}_{\msc{\bf M}}}
\newcommand{\Chg}[1]{\mbox{Ch}^{#1}_{\msc{\bf G}}}

\newcommand{\sn}{\msc{\bf n}}

\newcommand{\tr}{\mbox{Tr}}
\newcommand{\mod}{\mbox{mod}}

\newcommand{\tpsi}{\tilde{\psi}}

\renewcommand{\Im}{\mbox{Im}}
\newcommand{\sIm}{\msc{Im}}
\renewcommand{\Re}{\mbox{Re}}

\newcommand{\nn}{\nonumber\\}

\newcommand {\eqn}[1]{(\ref{#1})}

\def\boxit#1{\vbox{\hrule\hbox{\vrule\kern8pt
\vbox{\hbox{\kern8pt}\hbox{\vbox{#1}}\hbox{\kern8pt}}
\kern8pt\vrule}\hrule}}
\def\mathboxit#1{\vbox{\hrule\hbox{\vrule\kern8pt\vbox{\kern8pt
\hbox{$\displaystyle #1$}\kern8pt}\kern8pt\vrule}\hrule}}

\makeatletter
\@addtoreset{equation}{section}
\def\theequation{\thesection.\arabic{equation}}
\makeatother

\begin{document}

\begin{titlepage}
 \
 \renewcommand{\thefootnote}{\fnsymbol{footnote}}
 \font\csc=cmcsc10 scaled\magstep1
 {\baselineskip=14pt
 \rightline{
 \vbox{\hbox{hep-th/0403193}
       \hbox{UT-04-12}
       }}}

 \baselineskip=20pt
\vskip 2cm
 
\begin{center}
 \centerline{\huge  $SL(2;\br)/U(1)$ Supercoset and Elliptic Genera}

\vskip 5mm

{\huge of Non-compact Calabi-Yau Manifolds} 

 \vskip 2cm
\noindent{ \large Tohru Eguchi  and Yuji Sugawara} \\
{\sf eguchi@hep-th.phys.s.u-tokyo.ac.jp~,~
sugawara@hep-th.phys.s.u-tokyo.ac.jp}
\bigskip

 \vskip .6 truecm
 {\baselineskip=15pt
 {\it Department of Physics,  Faculty of Science, \\
  University of Tokyo \\
  Hongo 7-3-1, Bunkyo-ku, Tokyo 113-0033, Japan}
 }
\end{center}

\bigskip

\begin{abstract}
We first discuss the relationship between the $SL(2;\br)/U(1)$ supercoset 
and $\cN=2$ Liouville theory and make a precise correspondence between 
their representations. We shall show that the discrete unitary 
representations of $SL(2;\br)/U(1)$ theory 
correspond exactly to those massless
representations of $\cN=2$ Liouville theory which are closed under
modular transformations and studied in our previous work \cite{ES-L}.

It is known that toroidal partition functions of $SL(2;\br)/U(1)$ theory 
(2D Black Hole) contain two parts, continuous 
and discrete representations. The contribution of continuous representations
is proportional to the space-time volume and is divergent in the
infinite-volume limit while the part of discrete representations 
is volume-independent.

In order to see clearly the contribution of discrete representations
we consider elliptic genus which projects out the contributions of
continuous representations: making use of the $SL(2;\br)/U(1)$,  
we compute elliptic genera for various non-compact space-times such as
the conifold, ALE spaces, Calabi-Yau 3-folds with $A_n$ singularities etc.
We find that these elliptic genera in general have a complex modular property
and are not Jacobi forms as opposed to 
the cases of compact Calabi-Yau manifolds.

\end{abstract}

\vfill

\setcounter{footnote}{0}
\renewcommand{\thefootnote}{\arabic{footnote}}
\end{titlepage}
\baselineskip 18pt


\section{Introduction}

~

Study of superstring vacua in irrational superconformal theories
has been a challenging problem. These theories describe 
superstrings propagating in non-compact curved space-time
often developing isolated singularities. Applications
of the study of this subject include the string theory 
on Calabi-Yau singularities and its holographic description, 
or the little string theory on NS5-branes in the T-dual picture
\cite{GV,OV,ABKS,GKP,GK,Pelc,ES1,Mizoguchi,Mizoguchi2,
Yamaguchi,NN,HK2,Murthy}.   
The analysis is also important for backgrounds with non-trivial
time dependence studied in the context of time-like Liouville 
theory \cite{GS2,ST,Schomerus}. 
Much of the efforts of understanding irrational (super)conformal theories
have centered around the study of Liouville field theory. 
See, for instance, \cite{Nakayama} for 
a recent review including a detailed list of literature.

One of the outstanding difficulties in these non-compact models
is the coexistence of both discrete and continuous spectra of primary
fields.
Characters of these representations mix in a non-trivial manner 
under modular transformations in contrast with 
rational theories. 

As an example, let us recall  
the superstring vacua of the type 
\begin{eqnarray}
    \mbox{Minkowski space-time}\, \otimes \, \mbox{$\cN=2$ minimal}\,
      \otimes \, \mbox{$\cN=2$ Liouville} ~.  \nonumber
\end{eqnarray}
Then, we had the following puzzle \footnote
   {This issue was first raised in \cite{Mizoguchi2}.}; 
\begin{enumerate}
 \item These backgrounds correspond to isolated singularities
      in Calabi-Yau spaces \cite{GV,OV,ABKS,GKP}, 
      and one expects the existence of massless 
      excitations describing the deformation of the
      singularities in the string spectra.
 \item However, the modular invariant partition functions 
       for such superconformal systems (in the infinite-volume limit) 
       contain only contributions from continuous representations 
       which possess the mass gap
       and no discrete representations corresponding to 
       massless excitations occur, as was studied for instance in \cite{ES1}. 
       
In our previous work \cite{ES-L} we have partially resolved this puzzle by
showing that 
\item Discrete massless modes representing non-trivial cycles 
appear in the {\it open} string channels of cylinder amplitudes
associated with supersymmetric boundary states.
We have obtained 
candidate Cardy boundary states which correspond to ZZ 
and FZZT branes in the bosonic Liouville theory \cite{FZZ,Teschner,ZZ}. 
See also \cite{ASY}. 
\end{enumerate}

In this paper in order to further study these problems, we analyze 
the $SL(2;\br)_k/U(1)$ Kazama-Suzuki model
\cite{KS}, which is known to be T-dual (mirror) 
to the $\cN=2$ Liouville theory \cite{FZZ2,GK,HK1,Tong}. 
After identifying the $SL(2;\br)_k/U(1)$ conformal blocks 
(branching functions) with the $\cN=2$ characters, 
we perform  the character expansion of the toroidal partition function of
the Kazama-Suzuki model in order to study its closed string spectrum. 
We will follow the analysis given in \cite{HPT} (see also
\cite{MOS,IKP}) 
of the partition function of the bosonic 
$SL(2;\br)/U(1)$-model \cite{GawK,Gaw}. It is well-known that the 
gauged WZW model for $SL(2;\br)/U(1)$ describes the geometry of 
a 2-dimensional black hole \cite{2DBH}.

We find an infra-red divergence in the partition function of the theory
corresponding to the infinite volume $V$ of the geometry of 2D black hole.
In the limit $V \,\rightarrow \, \infty$ the partition function becomes 
simply the diagonal sum of continuous representations.
When we suitably regulate the IR divergence, however, we find a non-trivial 
weight function for the continuous series and also a set of
discrete representations with spin $1/2\le j\le (k+1)/2$. We find that these 
discrete series are exactly the
massless representations in ${\cal N}=2$ Liouville theory which are
closed under modular transformations
\cite{ES-L}.

Discrete states are localized around the tip of the cigar in 2D black 
hole \cite{DVV-BH}.
Thus they are suppressed in the infinite-volume limit as compared 
with the continuous representations.

In order to see clearly identify the contributions of the discrete states
not buried under the continuous representations,
we propose to study the elliptic genus 
of the theory \cite{Witten-E} to which the continuous
representations do not contribute.
We consider coupled superconformal systems: 
$SL(2;\br)_k/U(1) \otimes \cM$, 
where $\cM$ are various $\cN=2$ RCFT's.
By taking suitable choices for $\cM$ we have computed elliptic genera
for the conifold, ALE spaces, Calabi-Yau 3-folds with $A_n$ singularities etc.
It turns out that 
in general elliptic genera have a complicated modular property
and are not Jacobi forms as in the case of compact Calabi-Yau manifolds.
They are instead expressed in terms of the Appell function
which features in the study of higher rank vector bundles over elliptic curves \cite{Pol,STT}.

~


While preparing this manuscript, we became aware of 
an interesting paper \cite{IPT} on the e-print Archive, where the authors studied  
the modular properties of the extended characters in the bosonic 
$SL(2;\br)/U(1)$-model defined in the similar manner as in \cite{ES-L},
and there are some overlaps with the present work.

~

\newpage
\section{Toroidal Partition Function for 
the $SL(2;\br)/U(1)$ Kazama-Suzuki Model}

~

\subsection{Preliminaries}

~

The Kazama-Suzuki supercoset model \cite{KS} for $SL(2;\br)_k/U(1)$ 
is defined as the coset CFT
\begin{eqnarray}
\frac{SL(2;\br)_{\kappa}\times SO(2)_1}{U(1)_{-(\kappa-2)}}~,
\end{eqnarray}
which is an $\cN=2$ SCFT with the central charge and level \footnote
 {Throughout this paper we denote the level of super $SL(2;\br)$ as $k$, 
   and the level of bosonic $SL(2;\br)$ as $\kappa=k+2$.}
\begin{eqnarray}
\hat{c}\equiv \frac{c}{3}=1+\frac{2}{k}~, ~~~ k\equiv \kappa-2~.
\end{eqnarray}
More explicitly, the total world-sheet action 
is written as
\begin{eqnarray}
 S(g,A,\psi^{\pm}, \tpsi^{\pm}) &=& \kappa S_{\msc{gWZW}}(g,A)
 + S_{\psi}(\psi^{\pm}, \tpsi^{\pm}, A)~, 
\label{total action}\\
\kappa S_{\msc{gWZW}} (g,A) &=& \kappa S^{SL(2;\bsr)}_{\msc{WZW}} (g)
+ \frac{\kappa}{\pi}\int_{\Sigma}d^2z\, 
\left\{\tr \left(\frac{\sigma_2}{2}g^{-1}\partial_{\bar{z}}g\right)A_z
+ \tr \left(\frac{\sigma_2}{2} \partial_z g g^{-1}\right)A_{\bar{z}}    
\right. \nn
&& \hspace{2cm} \left. 
+ \tr\left(\frac{\sigma_2}{2}g\frac{\sigma_2}{2} g^{-1} \right)
   A_{\bar{z}}A_z + \frac{1}{2} A_{\bar{z}}A_z  \right\} ~, 
\label{gWZW action} \\
S^{SL(2;\bsr)}_{\msc{WZW}} (g) &=& -\frac{1}{8\pi} \int_{\Sigma} d^2z\,
\tr \left(\partial_{\al}g^{-1}\partial_{\al}g\right) +
\frac{i}{12\pi} \int_B \,\tr\left((g^{-1}dg)^3\right) ~,
\label{SL(2) WZW action} \\
 S_{\psi}(\psi^{\pm}, \tpsi^{\pm}, A)& =& 
\frac{1}{2\pi}\int d^2z\, \left\{
\psi^+(\partial_{\bar{z}}-A_{\bar{z}}) \psi^-
+\psi^-(\partial_{\bar{z}}+A_{\bar{z}}) \psi^+ \right. \nn 
&& \hspace{1in} \left. +\tpsi^+(\partial_{z}-A_{z}) \tpsi^-
+\tpsi^-(\partial_{z}+A_{z}) \tpsi^+
\right\} ~,
\label{fermion action}
\end{eqnarray}
where the complex fermions $\psi^{\pm}$ (and $\tpsi^{\pm}$)
have charge $\pm 1$ with respect to  the $U(1)$-gauge group. 
The bosonic part $\kappa S_{\msc{gWZW}}(g,A)$ is 
the gauged WZW model for the coset $SL(2;\br)_{\kappa}/U(1)_A$ 
\cite{KarS,GawK}, 
where $U(1)_A$ indicates the gauging 
the axial $U(1)$-symmetry; $g\,\rightarrow\, \Omega g \Omega$,
$\Omega(z,\bar{z}) = e^{iu(z,\bar{z}) \sigma_2}$ ($u(z,\bar{z})\in
\br$, $\sigma_2$ is the Pauli matrix.)  
It is well-known that this model describes string theory on 2D Euclidean 
black-hole with the cigar geometry \cite{2DBH}. 
The WZW action $\kappa S^{SL(2;\bsr)}_{\msc{WZW}} (g)$ is 
formally equal to  $-\kappa S^{SU(2)}_{\msc{WZW}}(g)$, 
and has a negative signature in $i\sigma_2$-direction.
Since we have $H^3(SL(2;\br))=0$,   
the action $\kappa S^{SL(2;\bsr)}_{\msc{WZW}} (g)$
can be rewritten as a purely two dimensional form and the level $\kappa$  
need not be an integer.

The chiral currents 
\begin{eqnarray}
&& j^A(z) = \kappa \tr \left(T^A\partial_z g g^{-1}\right)~, 
~~~ \tilde{j}^A(\bar{z}) = -\kappa \tr 
\left(T^A g^{-1}\partial_{\bar{z}} g\right) ~, 
\label{WZW currents} \\
&& T^3 = \frac{1}{2}\sigma_2~,~~ T^{\pm} = \pm
\frac{1}{2}\left(\sigma_3\pm i\sigma_1\right) 
\end{eqnarray}
satisfy the affine $\widehat{SL}(2;\br)_{\kappa}$ current algebra
(we write the left-mover only);
\begin{equation}
\left\{
\begin{array}{lll}
j^3(z)j^3(0) &\sim& \dsp -\frac{\kappa/2}{z^2} \\
j^3(z)j^{\pm}(0)&\sim& \dsp \frac{\pm 1}{z}j^{\pm}(0) \\
j^+(z)j^-(0) &\sim& \dsp \frac{\kappa}{z^2}-\frac{2}{z}j^3(0)
\end{array}
\right.
\label{SL 2}
\end{equation}
and the pair of free fermions $\psi^+$, $\psi^-$ satisfy 
the OPE's $\dsp \psi^+(z)\psi^-(0) \sim 1/z$,
$\psi^{\pm}(z)\psi^{\pm}(0)\sim 0$. 
The explicit realization of $\cN=2$ SCA is given by  
\begin{eqnarray}
&&T(z)= \frac{1}{k} \eta_{AB}j^Aj^B + \frac{1}{k} J^3 J^3 
- \frac{1}{2}(\psi^+\partial \psi^- - \partial \psi^+ \psi^-)~,~~~
(\eta_{AB}= \mbox{diag}(1,1,-1))~, \nn
&& J = \psi^+\psi^- + \frac{2}{k}J^3~, ~~~
 G^{\pm} = \frac{1}{\sqrt{k}} \psi^{\pm}j^{\mp}~, 
\label{N=2 SCA}
\end{eqnarray}
where we set $J^3 \equiv j^3 + \psi^+\psi^-$, which is 
the unique $U(1)$-current commuting with all the generators of 
$\cN=2$ SCA and hence defines the denominator of the 
$SL(2;\br)_k/U(1)$-supercoset.

To close this preliminary subsection, 
we summarize irreducible representations of $\widehat{SL}(2;\br)_{\kappa}$
current algebra.  We concentrate on the representations 
with conformal weights bounded from below\footnote
  {We have more general representations constructed by the spectral
flows;
$$
j^3_m~\rightarrow~j^3_m-\frac{\kappa}{2}n\delta_{m,0}~,~~~
j^{\pm}_m~\rightarrow~j^{\pm}_{m\pm n}~,~~~
L_m~\rightarrow~L_m+nj^3_m-\frac{\kappa}{4}n^2\delta_{m,0}~.
$$
However, they generically have unbounded spectra of conformal weights.  
}
and corresponding  to unitary representations of the zero-mode subalgebra
$\{j^3_0,\,j^{\pm}_0\}$.
The parameter $j$ is related to the conformal weight of vacuum 
states by the well-known formula
\begin{eqnarray}
h=-\frac{j(j-1)}{\kappa-2}~,
\label{h SL(2)}
\end{eqnarray}
in all cases.
\begin{description}
 \item[1. continuous series  :]  $\hC_{p,\al}$  
($\dsp j=\frac{1}{2}+ip$, $p\in \br_{\geq 0}$, $0\leq \al <1$)

They are non-degenerate representations and all the states lie above 
the mass gap $\dsp h\geq \frac{1}{4(\kappa-2)}$.
The vacua have the $j^3_0$-spectrum; $j^3_0=\al+n$, $n\in \bz$.
The character formula is simply given by ($q\equiv e^{2\pi i \tau}$, 
$y\equiv e^{2\pi i u}$);
\begin{eqnarray}
\chi_{p,\al}(\tau,u)=\frac{q^{\frac{p^2}{\kappa-2}}}{\eta(\tau)^3}
  \, \sum_{n\in\bsz}\,y^{n+\al}~.
\label{character C}
\end{eqnarray}
In the following arguments we often use formal identities such as
$$
 \frac{1}{1-y} + \frac{y^{-1}}{1-y^{-1}} = \sum_{n\in\bsz}\,y^n
    = \sum_{m\in \bsz}\,\delta(u+m)~,~~~(z\in \br).
$$
 For a more rigorous treatment, one may consider 
the ``regularized characters'' $\chi_{\xi}(\tau,u;\ep)$ defined by
replacing $e^{2\pi i (n+\al) u}$ 
with $e^{2\pi i (n+\al) u} e^{-|n|\ep}$ ($\ep>0$, $n\in \bz$, 
$0\leq \al <1$). 
\item[2. discrete series :] $\hD_j^{\pm}$ ($j \in \br_{>0}$)

The superscript $+$ indicates that these are 
the lowest weight representations and
$-$ does the highest weight ones. 
The spin parameter $j$ is allowed to be continuous despite the
name ``discrete series''
(while, in the $SL(2;\br)/U(1)$-coset theory $j$ takes discrete values).
The vacua have 
the $j^3_0$-spectrum; $j^3_0=\pm (j+n)$, $n\in \bz_{\geq 0}$
for $\hD^{\pm}_j$ respectively.
The character formula is written as 
\begin{eqnarray}
\chi^{\pm}_{j}(\tau,u)=\pm i 
\frac{q^{-\frac{1}{\kappa-2}\left(j-\frac{1}{2}\right)^2} 
y^{\pm\left(j-\frac{1}{2}\right)}}{\th_1(\tau,u)}~. 
\label{character Dj}
\end{eqnarray}
\item[3. identity representation :] ~

This is the representation generated 
by the vacuum $h=j^3_0=0$ corresponding to the
identity operator. 
Since this vacuum  is both highest and lowest weight, 
the character formula becomes
\begin{eqnarray}
\chi_0(\tau,u)=i\frac{q^{-\frac{1}{4(\kappa-2)}}
y^{-1/2}(1-y)}{\th_1(\tau,u)} \equiv
\frac{q^{-\frac{\kappa}{8(\kappa-2)}}}
{\prod_{n=1}^{\infty}(1-q^n)(1-yq^n)(1-y^{-1}q^n)}.
\end{eqnarray}
\item[4. complementary representations] : $\hE_{j,\al}$
$\dsp 
(0<j<\frac{1}{2}, ~ \left|j-\frac{1}{2}\right| <\left|\al-\frac{1}{2}\right|,~
0\leq \al <1)$

These are non-degenerate representations 
below the mass gap. The vacua again have 
the $j^3_0$-spectrum;
$j^3_0= \al+n$, ($n\in \bz$). The character formula has the same form 
as the continuous series. The range of $j$ comes from the unitarity of 
zero-mode subalgebra. At the ``boundary'' of this range 
$j=0$, $\al=0$, the representation with $h=0$ becomes  
reducible and is decomposed as 
\begin{eqnarray}
\hE_{j=0,\al=0} \cong (\mbox{identity rep.}) \oplus \hD^+_{j=1} 
\oplus \hD^-_{j=1}~.
\label{decomposition h=0}
\end{eqnarray}

\end{description}

~

\subsection{Branching Functions}

~

We start our analysis by identifying the conformal blocks 
for the toroidal partition function in the $SL(2;\br)_k/U(1)$ 
Kazama-Suzuki model \cite{KS}. 
Although this task has been already 
implicitly carried out in \cite{DPL}\footnote
    {It is an easy exercise to show the equivalence of
     the ``non-compact parafermion''  
     approach (with a compact boson added) used in \cite{DPL} 
     with the $SL(2;\br)/U(1)$ Kazama-Suzuki model. We can thus 
     associate all the branching functions of the Kazama-Suzuki model
     with unitary irreducible representations of $\cN=2$ SCA based 
     on the analysis given in \cite{DPL}. See also \cite{FST}.
     }, 
it is helpful 
for our later analysis to provide the explicit formulas of 
conformal blocks. It turns out that these blocks are identified with 
the irreducible characters of $\cN=2$ SCA. This fact 
will be the simplest evidence for T-duality with 
the $\cN=2$ Liouville theory.

We focus on the NS-sector and the formulas for other spin structures 
are obtained by using the $1/2$-spectral flow. 
According to the standard treatment of coset CFT,  
the conformal blocks are defined as the following branching
functions;
\begin{eqnarray}
\chi_{\xi}(\tau,u)\frac{\th_3(\tau,v)}{\eta(\tau)}
=\sum_{m}\chi^{(\sNS)}_{\xi,m}(\tau,z)
\frac{q^{-\frac{m^2}{k}}e^{2\pi i mw}}{\eta(\tau)}~.
\label{branching 0}
\end{eqnarray}
where $\xi$ indicates each of the irreducible representations of 
$\widehat{SL}(2;\br)_{\kappa}$ classified above.  
The angle variables $u$, $v$, $z$, $w$ are associated to the
$U(1)$-currents $j^3$, $\psi^+\psi^-$, $J$, $J^3$ respectively, 
and we can easily read off the relations among them as 
\begin{eqnarray}
u=\frac{2}{k}z+w~,~~~v=u+z\equiv \frac{k+2}{k}z+w~,
\label{u v}
\end{eqnarray}
from the definitions $J^3=j^3+\psi^+\psi^-$, 
$\dsp J=\psi^+\psi^-+ \frac{2}{k}J^3$.  
The summation of $m$ runs over the possible $J^3_0$-spectrum 
of representation $\xi$ (tensored with the free fermion system
$\psi^{\pm}$)

The branching functions for continuous series (and the complementary 
representations) are easily obtained due to the absence of 
null states;
\begin{eqnarray}
&& \mathboxit
{
\chic^{(\sNS)}_{\,p,m}(\tau,z)= q^{\frac{p^2+m^2}{k}}\, 
e^{2\pi i \frac{2m}{k}z} \,\frac{\th_3(\tau,z)}{\eta(\tau)^3}
\equiv \ch{(\sNS)}{}
\left(h=\frac{p^2}{k}+\frac{1}{4k}+\frac{m^2}{k}, 
Q=\frac{2m}{k};\tau,z\right)
} \nn
&&
\label{branching C}
\end{eqnarray}
Here $\ch{(\sNS)}{}$ denotes 
the $\cN=2$ irreducible character for the massive (non-degenerate) 
representation \eqn{massive character}.
We sometimes allow the pure imaginary values of $p$ corresponding to 
the branching functions for the complementary representations 
$\hE_{j,\al}$ ($\dsp j= \frac{1}{2}+ip$). 
The unitarity condition \cite{BFK} is derived from those for  
(the zero-mode parts of) $\hC_{p,\al}$, $\hE_{j=\frac{1}{2}+ip, \al}$;
\begin{eqnarray}
p^2+\left(\al-\frac{1}{2}\right)^2 \geq 0~,~~~ \al \equiv m ~(\mod\, 1)~,
~~ 0\leq \al <1~.
\label{unitarity cont}
\end{eqnarray}


Derivation of the branching functions for the discrete series 
is more non-trivial.
We will focus on $\hD_j^+$ because $\hD_j^-$ is obtained by the spectral flow 
\begin{eqnarray}
q^{-\frac{\kappa}{4}} e^{-2\pi i \frac{\kappa}{2} u}\, \chi^+_j(\tau,u+\tau)
=\chi^-_{\frac{\kappa}{2}-j}(\tau,u)~.
\label{D + -}
\end{eqnarray}
The desired branching function is expressed as   
\begin{eqnarray}
\chid_{\,j,j+n}^{(\sNS)}(\tau,z) &\equiv &
q^{\frac{(j+n)^2}{k}}\,
\int_0^1dw\, \chi_j^+(\tau,u) \th_3(\tau,v) e^{-2\pi i (j+n) w} ~. 
\label{branching D 0}
\end{eqnarray}
We first consider the following shifts of angular variables in
\eqn{branching D 0} (or \eqn{branching 0});
\begin{eqnarray}
z~\longmapsto ~ z+n\tau~,~~~ w~\longmapsto ~ w- \frac{2}{k}n\tau~,
~~~({}^{\forall}n\in \bz)
\end{eqnarray}
which leaves $u$ invariant and causes $v\,\longmapsto\,v+n\tau$.
Using the property $\th_3(\tau,v+n\tau)=
q^{-\frac{n^2}{2}}e^{-2\pi in v}\th_3(\tau,v)$, we find the relation
\begin{eqnarray}
q^{\frac{\hat{c}}{2}n^2}e^{2\pi i \hat{c} n z}\,
\chid_{\,j,j}^{(\sNS)}(\tau,z+n\tau)= 
\chid_{\,j,j+n}^{(\sNS)}(\tau,z)~,~~~(\hat{c}=1+\frac{2}{k})~.
\label{spectral flow branching}
\end{eqnarray}
Therefore, it is enough to calculate  $\chid^{(\sNS)}_{\,j,j}(\tau,z)$. 
The easiest  way to evaluate it is to use of the  
character relation
\begin{eqnarray}
\chi_j^+(\tau,u)+q^{\frac{\kappa}{4}-j}e^{2\pi i \frac{\kappa-2}{2}u}
\chi_{\frac{\kappa}{2}-j}^-(\tau,u) = \chi_{j,\al=j}(\tau,u)
~.
\label{character relation 1}
\end{eqnarray}
R.H.S is the character of complementary 
representation $\hE_{j,\al=j}$.
 
We act by $\dsp q^{\frac{j^2}{k}} \,\int_0^1dw\, 
\th_3(\tau,v) e^{-2\pi i jw}$ on both sides of 
\eqn{character relation 1}.
We also need the branching relation for $\hD^-_{\frac{\kappa}{2}-j}$;
\begin{eqnarray}
\chid_{\,j,j+n}^{(\sNS)}(\tau,z) &= &
q^{\frac{\left(j-\frac{k}{2}+n\right)^2}{k}}\,
\int_0^1dw\, \chi_{\frac{\kappa}{2}-j}^-(\tau,u) \th_3(\tau,v) 
e^{-2\pi i \left(j-\frac{k}{2}+n\right) w}~,
\label{branching D 2}
\end{eqnarray}
which is derived from \eqn{D + -} by making a shift 
$w\, \longmapsto\,w+\tau$ with keeping $z$ (that is, 
$u\, \longmapsto\, u+\tau$, $v\,\longmapsto\,v+\tau$)
in \eqn{branching D 0}. 
We then obtain 
\begin{eqnarray}
(1+e^{2\pi i z}q^{1/2}) \chid^{(\sNS)}_{\,j,j}(\tau,z) = 
\chic^{(\sNS)}_{\,j,m=j}(\tau,z) \equiv 
q^{\frac{j}{k}-\frac{1}{4k}}e^{2\pi i \frac{2j}{k}z}\, 
\frac{\th_3(\tau,z)}{\eta(\tau)^3}~.
\end{eqnarray}
This leads to the formula 
\begin{eqnarray}
\chid_{\,j,j}^{(\sNS)}(\tau,z) = 
\frac{q^{\frac{j}{k}-\frac{1}{4k}} e^{2\pi i \frac{2j}{k}z}}
{1+e^{2\pi i z}q^{1/2}}
\, \frac{\th_3(\tau,z)}{\eta(\tau)^3} 
\equiv  \chm{(\sNS)}(Q=2j/k;\tau,z)~.
\label{branching D}
\end{eqnarray}
Here $\chm{(\sNS)}(Q;\tau,z)$ is the massless matter character
of $\cN=2$ SCA for the chiral primary states $h=Q/2>0$ 
\eqn{massless character 1}.
Since \eqn{spectral flow branching} is the proper relation
for spectral flow of $\cN=2$ SCA (see \eqn{flowed character}),
we can identify $\chid_{\,j,j+n}^{(\sNS)}(\tau,z)$ with 
the flowed massless matter character \eqn{flowed character};
\begin{eqnarray}
&&
\mathboxit
{
\chid_{\,j,j+n}^{(\sNS)}(\tau,z) =
\chm{(\sNS)}(Q=2j/k,n;\tau,z)
\equiv  \frac{q^{\frac{j+n^2+2nj}{k}-\frac{1}{4k}} 
e^{2\pi i \frac{2(j+n)}{k}z}}{1+e^{2\pi i z}q^{n+1/2}}\,
\frac{\th_3(\tau,z)}{\eta(\tau)^3}~, ~~~ ({}^{\forall}n\in\bz)
} \nn
&&
\label{branching D flowed}
\end{eqnarray}
One may check directly the validity of our branching relation
\begin{eqnarray}
\chi^+_j(\tau,u)\frac{\th_3(\tau,v)}{\eta(\tau)}=
\sum_{n\in\bsz}\frac{q^{\frac{j+n^2+2nj}{k}-\frac{1}{4k}} 
e^{2\pi i \frac{2(j+n)}{k}z}}{1+e^{2\pi i z}q^{n+1/2}}\,
\frac{\th_3(\tau,z)}{\eta(\tau)^3} \cdot 
\frac{q^{-\frac{(j+n)^2}{k}}
e^{2\pi i (j+n)w}}{\eta(\tau)}~,
\label{branching check}
\end{eqnarray}
by comparing the residues at poles $e^{2\pi i u} q^m=1$ \,($m \in \bz$)
of both sides. 


It is useful to note:
\begin{itemize}
 \item $n\geq 0$: The vacuum state of $\chid^{(\sNS)}_{\,j,j+n}(\tau,z)$
is $(j_0^+)^n\ket{j,j}
\otimes \ket{0}_{\psi}$,
which possesses the quantum numbers
\begin{eqnarray}
h= \frac{2j\left(n+\frac{1}{2}\right)+n^2}{k}~,~~~ Q= \frac{2(j+n)}{k}~,
\label{vacuum 1}
\end{eqnarray}
 \item $n<0$: The vacuum state is $(j^-_{-1})^{|n|-1}\ket{j,j} \otimes 
  \psi^-_{-1/2}\ket{0}_{\psi}$,
which has
\begin{eqnarray}
h= \frac{-\left(k-2j\right)\left(n+\frac{1}{2}\right)+n^2}{k}~,
~~~ Q= \frac{2(j+n)}{k}-1~. 
\label{vacuum 2}
\end{eqnarray}
\end{itemize}
Especially, $\chid_{\,j,j-1}^{(\sNS)}(\tau,z)$
is the character of anti-chiral primary with 
$\dsp h= -\frac{Q}{2}=\frac{1}{k}\left(\frac{k+2}{2}-j\right)$.
As is proved in \cite{DPL}, the unitarity bound for the Casimir
parameter $j$ is given as 
\begin{eqnarray}
 0 < j < \frac{\kappa}{2} \left(\equiv \frac{k+2}{2}\right)~,
\label{unitarity j}
\end{eqnarray}
which reproduces all the (spectrally flowed) massless matter 
representations of $\cN=2$ SCA
lying on the ``unitarity segments''  given in \cite{BFK}. 


Branching functions for the identity 
representation $\chid^{(\sNS)}_{\,j=0,m=0}(\tau,z)$ 
may be derived in a similar manner with the help of \eqn{decomposition h=0}.
As one may expect, it is given by the graviton representation $h=Q=0$ 
\eqn{massless character 2} of ${\cal N}=2$ SCA.
We find that 
\begin{equation}
\chid^{(\sNS)}_{\,0,0}(\tau,z) = \chg{(\sNS)}(\tau,z)=q^{-{1\over 4k}}{1-q\over (1+e^{2\pi iz}q^{1/2})
(1+e^{-2\pi iz}q^{1/2})}{\theta_3(\tau,z)\over \eta(\tau)^3}
\end{equation}
Spectral-flowed version is given by
\begin{eqnarray}
&& 
\mathboxit
{
\chid^{(\sNS)}_{\,0,n}(\tau,z) = \chg{(\sNS)}(n;\tau,z)
\equiv  q^{-\frac{1}{4k}} \frac{(1-q)q^{\frac{n^2}{k}+n-\frac{1}{2}}
  e^{2\pi i \left(\frac{2n}{k}+1\right)z}}
  {\left(1+e^{2\pi i z}q^{n+1/2}\right)\left(1+e^{2\pi i z}q^{n-1/2}\right)}\,
  \frac{\th_3(\tau,z)}{\eta(\tau)^3}
} \nn
&& 
\label{branching Id}
\end{eqnarray}
The corresponding vacua are slightly non-trivial;
\begin{itemize}
 \item $n=0$ : The vacuum is $\ket{0,0}\otimes \ket{0}_{\psi}$ with 
$h=Q=0$.
 \item $n\geq 1$ : The vacuum is $(j^+_{-1})^{n-1}\ket{0,0} \otimes 
\psi^+_{-1/2}\ket{0}_{\psi}$, which has the quantum numbers 
 \begin{eqnarray}
  h= \frac{n^2}{k}+n-\frac{1}{2}~,~~~ Q= \frac{2n}{k}+1~. 
 \end{eqnarray}
 \item $n\leq -1$ : The vacuum is $(j^-_{-1})^{|n|-1}\ket{0,0} \otimes 
\psi^-_{-1/2}\ket{0}_{\psi}$, which has the quantum numbers
 \begin{eqnarray}
  h= \frac{n^2}{k}-n-\frac{1}{2}~,~~~ Q= \frac{2n}{k}-1~. 
 \end{eqnarray}
\end{itemize}


We finally introduce the branching functions of other spin structures 
to fix the convention in this paper. 
Let $\chi^{(\sNS)}_{*,m}(\tau,z)$ be the abbreviated notations of
the branching functions considered above ($m=J^3_0$).   We define
\begin{eqnarray}
&& \chi^{(\stNS)}_{*,m}(\tau,z) \equiv e^{-i\pi \frac{2m}{k}}\,
\chi^{(\sNS)}_{*,m}\left(\tau,z+\frac{1}{2}\right)~, \nn
&& \chi^{(\sR)}_{*,m+\frac{1}{2}}(\tau,z) \equiv 
  q^{\frac{k+2}{8k}} e^{i\pi z \frac{k+2}{k}}\,
 \chi^{(\sNS)}_{*,m}\left(\tau,z+\frac{\tau}{2}\right)~, \nn
&& \chi^{(\stR)}_{*,m+\frac{1}{2}}(\tau,z) \equiv 
  e^{-i\pi \frac{2m}{k}} q^{\frac{k+2}{8k}} e^{i\pi z \frac{k+2}{k}}\,
 \chi^{(\sNS)}_{*,m}\left(\tau,z+\frac{\tau}{2}+\frac{1}{2}\right)~.
\label{branching os}  
\end{eqnarray}

~


\subsection{Toroidal Partition Function}

~

Let us analyze the toroidal partition function 
of $SL(2;\br)_k/U(1)$ Kazama-Suzuki model.
It can be evaluated by the path-integral approach as described
in \cite{GawK,Gaw} for the bosonic 2D black-hole model 
\cite{2DBH}. Here we present only the result and leave 
the detailed calculations to Appendix C.    
For the NS sector of the theory, we obtain 
\begin{eqnarray}
Z^{(\sNS)}(\tau) &=& \int \cD\lb g, A, \psi^{\pm}, \tpsi^{\pm}\rb\,
e^{-\kappa S_{\msc{gWZW}}(g,A) - S_{\psi}(\psi^{\pm},\tpsi^{\pm}, A)}~\nn
&=& C \int_0^1ds_1 \int_0^1ds_2\, 
\frac{\left|\th_3(\tau,s_1\tau-s_2)\right|^2}
{\left|\th_1(\tau,s_1\tau-s_2)\right|^2} \,
\sum_{w,m\in\bsz}\, \exp \left(-\frac{\pi k}{\tau_2}
\left|(w+s_1)\tau-(m+s_2)\right|^2\right)~, \nn
&& 
\label{part fn}
\end{eqnarray}
where $C$ is a normalization constant to be fixed later.
The partition functions for other spin structures are 
obtained by simply replacing $\th_3(\tau,s_1\tau-s_2)$ 
with $\th_{\lb \sigma \rb}(\tau,s_1\tau-s_2)$,
defined as 
$\th_{\lb \sNS \rb} = \th_3$,
$\th_{\lb \stNS \rb} = \th_4$,
$\th_{\lb \sR \rb} = \th_2$ and 
$\th_{\lb \stR \rb} = i\th_1$.
Assuming the diagonal modular invariant for spin 
structures, we obtain the partition function
\begin{eqnarray}
Z(\tau) = \frac{1}{2}\sum_{\sigma}\, Z^{(\sigma)}(\tau)~.
\label{part fn 2}
\end{eqnarray}
Here $u\equiv s_1\tau-s_2$ ($0\leq s_1,s_2 \le 1$) is the modulus 
of gauge field $A$. One can view the sum over $m,n$ as summing over the
momentum and winding modes of a compact boson $Y$ which parameterizes the 
2-dimensional gauge field $A$. $Y$ has a 
radius $\sqrt{2k}$ which is the size of the asymptotic 
circle far from the tip of the cigar. 
With the canonical normalization $Y(z)Y(0)\sim -\ln z$ for the field $Y$, 
total anomaly free current 
defining the BRST charge (see \cite{KarS}) is given as 
\begin{eqnarray}
J^3_{\msc{tot}} \equiv j^3+\psi^+\psi^- + \sqrt{\frac{k}{2}}i\partial Y~,~~~
\tJ^3_{\msc{tot}} \equiv \tj^3+\tpsi^+\tpsi^- - \sqrt{\frac{k}{2}}i\deebar Y~.
\label{total current}
\end{eqnarray}
Of course, these currents 
have no singular OPE's with the $\cN=2$ SCA generators 
\eqn{N=2 SCA}, assuring their  BRST-invariance.

The partition function \eqn{part fn 2} is modular invariant 
in a formal sense since the modulus integral 
$\dsp \int ds^1 ds^2$ is logarithmically divergent due to 
the double pole of $1/|\th_1(\tau,s_1\tau-s_2)|^2$.
The appearance of divergence is not surprising since the target space
has an infinite volume. 
The evaluation of modulus integral 
with a suitable IR cut-off in 
\eqn{part fn} is quite useful in determining the closed string spectrum, 
as shown in \cite{HPT} for the bosonic $SL(2;\br)/U(1)$ model. 
We turn to this analysis from now on.


~


\subsection{Expansion into Branching Functions}

~

We expand the toroidal partition function into 
branching functions of $SL(2;\br)_k/U(1)$ 
following the procedure of \cite{HPT}. 
(See also \cite{MOS,IKP}.) Although our result 
will be quite similar to the bosonic 
case \cite{HPT}, we will present our analysis
for the supersymmetric case for the sake of completeness.

We start with the partition function of NS sector \eqn{part fn}. 
Using the Poisson resummation formula, 
we can rewrite it as;
\begin{eqnarray}
&& Z^{(\sNS)}(\tau) = C \int_0^1ds_1\,\int_0^1ds_2\, 
\sqrt{\frac{\tau_2}{k}} \sum_{w,n\in \bsz}\, 
\frac{\left|\th_3(\tau,s_1\tau-s_2)\right|^2}
{\left|\th_1(\tau,s_1\tau-s_2)\right|^2} \,
e^{-2\pi \tau_2\left(\frac{n^2}{2k}+\frac{k}{2}(s_1+w)^2\right)
-2\pi i n \left((s_1+w)\tau_1-s_2\right)}~.\nn
&& \label{expansion 1}
\end{eqnarray}
The $s_2$-integral is easily performed since $s_2$ appears only linearly
in the exponent. 
The $q$-expansion of the theta function terms is 
expressed as the trace over the Hilbert space of 
various oscillators.
We introduce the oscillator levels $N$, $\tN$
and also the operators $l$, $\tl$ defined as 
\begin{eqnarray}
&& l 
\equiv  
\sharp \{j^+_n,\, \psi^+_r\}- \sharp \{j^-_n,\, \psi^-_r\}~, ~~~
\tl 
\equiv \sharp \{\tj^+_n,\, \tpsi^+_r\}- \sharp \{\tj^-_n,\, \tpsi^-_r\}~.
\end{eqnarray}
The relevant Hilbert space is 
\begin{eqnarray}
\cH^{\pm} \equiv \left\lb \cF_{SL(2)}^{\pm}\otimes \cF_{\psi}
\otimes \cF_Y \otimes \cF_{\msc{gh}}\right\rb_L \otimes
\left\lb \cF_{SL(2)}^{\pm} \otimes \cF_{\psi}
\otimes \cF_Y \otimes \cF_{\msc{gh}}\right\rb_R~,
\label{Fock space 1}
\end{eqnarray}
where
$\cF_*$ denotes the Fock space of each sector. 
Especially, in the $SL(2;\br)$-sector, $\cF_{SL(2)}^{+}$ and 
$\cF_{SL(2)}^{-}$ means respectively the ones associated to 
the lowest and highest weight representations of zero-modes: 
namely we have 
\begin{eqnarray}
\tr_{\cF_{SL(2)}^{\pm}} \left(q^{N-1/8} e^{2\pi i  u l}\right) = 
\frac{\pm i e^{\mp i \pi  u}}{\th_1(\tau,u)}~.
\end{eqnarray}
We thus obtain 
\begin{eqnarray}
&&Z^{(\sNS)}(\tau) 
= C \int_0^1ds_1 \,\sqrt{\frac{\tau_2}{k}} \sum_{w,n\in \bsz}\,
e^{-2\pi \tau_2\left(\frac{n^2}{2k}+\frac{kw^2}{2}+kws_1
+\frac{k}{2}s_1^2\right)}\, \nn
&& \hspace{3cm} \times  \tr_{\cH^{+}}\left( 
e^{-2\pi \tau_2 \left(N+\tN+(l+\tl+1)s_1 -\frac{1}{4}\right) 
+2\pi i \tau_1(N-\tN-nw)}
\right) ~, 
\label{expansion 2}
\end{eqnarray}
where the trace is constrained by the condition $l-\tl=n$ imposed by 
the $s_2$-integral.

The $s_1$-integral is non-trivial since $s_1$ appears quadratically in the exponent. 
Following \cite{MOS,HPT}, we linearize it 
by means of the Fourier transformation;
\begin{eqnarray}
\sqrt{k\tau_2} e^{-2\pi \tau_2 \frac{k}{2}s_1^2} =
\int_{-\infty}^{\infty} dc\, e^{-\frac{\pi}{k\tau_2}c^2 -2\pi i cs_1}~.
\end{eqnarray}
The $s_1$-integral is then  easy to compute  and gives 
\begin{eqnarray}
&&\int_{-\infty}^{\infty}dc\,\int_0^1ds_1\,
e^{-\frac{\pi}{k\tau_2}c^2 -2\pi \tau_2 
\left(\frac{n^2}{2k}+\frac{k}{2}w^2\right)
-2\pi s_1\left(ic+\tau_2(kw+l+\tl+1)\right)} \nn
&& ~~~ = 
\int_{-\infty}^{\infty} dc\, 
\frac{-e^{-\frac{\pi}{k\tau_2}c^2 -2\pi \tau_2 
\left(\frac{n^2}{2k}+\frac{k}{2}w^2\right)} }
{2\pi\left(ic +\tau_2(kw+l+\tl+1)\right)}\,
\left\{e^{-2\pi 
\left(ic+\tau_2(kw+l+\tl+1)\right)} - 1
\right\} \nn
&& ~~~ = -\frac{1}{2\pi} \int_{C_2}dp\, 
\frac{e^{-2\pi \tau_2\left(2\frac{p^2}{k}+
\frac{n^2}{2k}+\frac{k}{2}(w+1)^2
+l+\tl+1\right)}}
{ip+\frac{1}{2}\left(k(w+1)+l+\tl+1\right)}
+ \frac{1}{2\pi} \int_{C_1}dp\, 
\frac{e^{-2\pi \tau_2 \left(2\frac{p^2}{k}
+\frac{n^2}{2k}+\frac{k}{2}w^2\right)}}
{ip+\frac{1}{2}\left(kw+l+\tl+1\right)} ~. \nn
&&
\label{evaluation 1}
\end{eqnarray}
In the last line we set $c=2\tau_2 p-i\tau_2 k$ in the first term,
and set $c=2\tau_2 p$ in the second term. The integration contours 
are defined as $C_1\,:\, \Im\,p=0$, ~$C_2\,:\, \Im\,p=k/2$.

To proceed further it is useful to make use of the spectral
flow associated to the total currents \eqn{total current},
defined symbolically as\footnote
    {This is different from the standard spectral flow 
  of $\cN=2$ SCA, defined in the same way referring to 
   the $\cN=2$ $U(1)$-currents (see \eqn{spectral flow}).
   We note that the operators $\cU_n$ 
   preserve the total current \eqn{total current} 
  and hence make the BRST-charge
   invariant. $\cU_n$ also preserves the $\cN=2$ SCA generators
   \eqn{N=2 SCA}. 
}
\begin{eqnarray}
\cU_n\equiv e^{in \Phi(0,0)}~, ~~~ i\partial \Phi \equiv J^3_{\msc{tot}}~,
~~  i\bar{\partial} \Phi \equiv \tJ^3_{\msc{tot}}~, ~~~(n\in \bz)~.
\end{eqnarray}
It is easy to see that $\cU_1$ acts as  
\begin{eqnarray}
&&\cU_1^{-1} \,N \,\cU_1 = N +l +\frac{1}{2}~, 
~~\cU_1^{-1} \,\tN \,\cU_1 = \tN +\tl +\frac{1}{2}~, \nn
&&
\cU_1^{-1} \,l \,\cU_1 = l+1~, ~~~\cU_1^{-1}\, \tl\, \cU_1 = \tl+1 ~, ~~~
 \cU_1^{-1}\, w\, \cU_1 = w+ 1~, ~~~ \cU_1^{-1} \,n \,\cU_1 = n~,
\end{eqnarray}
and maps the Fock space $\cH^+$ to $\cH^-$. 
We thus find that
\begin{eqnarray}
&&\sum_{w,n\in\bsz}\,
e^{-2\pi \tau_2 \left(\frac{n^2}{2k}+\frac{k}{2}(w+1)^2\right)} \,
\tr_{\cH^+} \,\left( 
\frac{e^{-2\pi \tau_2\left(N+\tN+l+\tl+1-\frac{1}{4}\right)+
2\pi i \tau_1\left(N-\tN-nw\right)}}{ip + \frac{1}{2}\left(
k(w+1)+l+\tl+1\right)}
\right) \nn
&& ~~~= \sum_{w,n\in \bsz}\,
e^{-2\pi \tau_2 \left(\frac{n^2}{2k}+\frac{k}{2}w^2\right)} \,
\tr_{\cH^-} \,\left( 
\frac{e^{-2\pi \tau_2\left(N+\tN-\frac{1}{4}\right)+
2\pi i \tau_1\left(N-\tN-nw\right)}}{ip + \frac{1}{2}\left(
kw+l+\tl-1\right)}
\right)~.
\label{evaluation 2}
\end{eqnarray}
Substituting \eqn{evaluation 1} and \eqn{evaluation 2} to \eqn{expansion 2}, 
we can show ($\hat{c}\equiv (k+2)/k$)
\begin{eqnarray}
&& Z(\tau) = \frac{C}{2\pi k} \sum_{w,n\in \bsz}\,
\left\lb 
\int_{C_1}dp\, e^{-2\pi \tau_2\left(\frac{n^2}{2k}+\frac{kw^2}{2}
+2\frac{p^2+1/4}{k}- \frac{\hat{c}}{4}\right)} \,
\tr_{\cH^+}\, \left( \frac{e^{-2\pi \tau_2\left(N+\tN\right)
+2\pi i \tau_1 \left(N-\tN-nw\right)}}
{ip  + \frac{1}{2}\left(kw+l+\tl+1\right)}
\right)
\right. \nn
&& \hspace{1cm}
\left. -\int_{C_2}dp\,
e^{-2\pi \tau_2\left(\frac{n^2}{2k}+\frac{kw^2}{2}
+2\frac{p^2+1/4}{k}- \frac{\hat{c}}{4}\right)}\,
\tr_{\cH^-}\, \left( \frac{e^{-2\pi \tau_2\left(N+\tN\right)
+2\pi i \tau_1 \left(N-\tN-nw\right)}}
{ip  + \frac{1}{2}\left(kw+l+\tl-1\right)}
\right)\right\rb~.
\label{expansion 3}
\end{eqnarray}
As in \cite{MOS,HPT}, let us perform the contour deformation;
$C_2\,\rightarrow\, C_1$, which picks up extra contributions from 
simple poles within the range  $0\leq \Im\,p \leq k/2$.
The partition function is now divided into two parts;
\begin{eqnarray}
Z^{(\sNS)}(\tau)= \Zc^{(\sNS)}(\tau) + \Zd^{(\sNS)}(\tau)~,
\end{eqnarray}
where the first term includes the $p$-integration on the real axis ($C_1$)
and the second corresponds to the sum of pole residues.

The first term $\Zc^{(\sNS)}(\tau)$ is rewritten as 
\begin{eqnarray}
&& \Zc^{(\sNS)}(\tau) = \frac{C}{2\pi k} \sum_{w,n\in \bsz}\, 
\int_{-\infty}^{\infty}dp\, e^{-2\pi \tau_2\left(\frac{n^2}{2k}+\frac{kw^2}{2}
+2\frac{p^2+1/4}{k}- \frac{\hat{c}}{4}\right)} \, \left\lb
\tr_{\cH^+}
\, \left( \frac{e^{-2\pi \tau_2\left(N+\tN\right)
+2\pi i \tau_1 \left(N-\tN-nw\right)}}
{ip  + \frac{1}{2}\left(kw+l+\tl+1\right)}
\right) \right. \nn
&& \hspace{4cm}
\left. 
-\tr_{\cH^-}
\, \left( \frac{e^{-2\pi \tau_2\left(N+\tN\right)
+2\pi i \tau_1 \left(N-\tN-nw\right)}}
{ip  + \frac{1}{2}\left(kw+l+\tl-1\right)}
\right)
\right\rb 
\label{expansion c 1}
\end{eqnarray}
Since the factor $l+\tl$ appears only in the denominators,
the traces are logarithmically divergent when one 
sums over the states of the form 
$(j^+_0\tj^+_0)^r\ket{\psi}$ ($(j^-_0\tj^-_0)^r\ket{\psi}$) 
for $\tr_{\cH^+}$ ($\tr_{\cH^-}$).  
This divergence comes 
from the pole $s_1=s_2=0$ in \eqn{part fn}, that is,
the infinite volume effect. 
Since the exponent $\dsp \frac{n^2}{2k}+ \frac{kw^2}{2}+
2\frac{p^2+1/4}{k} -\frac{\hat{c}}{4}$ is the correct weights for  
the continuous representations, 
it is natural to expect that $\Zc^{(\sNS)}(\tau)$ can be expressed 
in a form
\begin{eqnarray}
&&\Zc^{(\sNS)}(\tau) = \frac{C}{k} \sum_{w,n\in \bsz}\,
\int_0^{\infty}dp\,\rho(p,w,n)\, \chic^{(\sNS)}_{\,p,m}(\tau,0)
\chic^{(\sNS)}_{\,p,\tm}(\bar{\tau},0)~,  \label{expansion c 2} \\
&& ~~~ m\equiv \frac{n-kw}{2}~,~~~\tm \equiv -\frac{n+kw}{2}~,
\label{m w n}
\end{eqnarray}
with a suitable spectral density $\rho(p,w,n)$.
Here $\chic^{(\sNS)}_{\,p,m}(\tau,0)$ is the branching function 
of continuous series \eqn{branching C} and is an
irreducible massive character of $\cN=2$ SCA.
Although there appear some ambiguities in regulating
the IR divergence, a candidate expression for $\rho(p,w,n)$
has been proposed in \cite{MOS,HPT}.
\begin{eqnarray}
\rho(p,w,n)=\frac{1}{2\pi}\,2 \log \,\ep + \frac{1}{2\pi i } 
\frac{d}{2dp}
\log\,\frac{\Gamma\left(-ip+\frac{1}{2}-m\right)
\Gamma\left(-ip+\frac{1}{2}+\tm\right)}{\Gamma\left(+ip+\frac{1}{2}-m\right)
\Gamma\left(+ip+\frac{1}{2}+\tm\right)}~,
\label{evaluation rho}
\end{eqnarray}
where $\ep>0$, $\ep\approx 0$ is the IR cut-off.
The first term in \eqn{evaluation rho} is interpreted as the 
volume factor. The second term has a non-trivial momentum dependence and 
is related to the reflection amplitudes of Liouville theory as is discussed in \cite{MOS}. 


On the other hand, the pole contributions yield  
the sum over the branching functions of discrete series \eqn{branching D}
(and \eqn{branching D flowed}).
We take the identification $\dsp j=-ip+\frac{1}{2}$
for the spin parameter $j$
so that we have $\dsp e^{-2\pi \tau_2\frac{p^2}{k}}
= e^{2\pi \tau_2\frac{1}{k}\left(j-\frac{1}{2}\right)^2}$
(with this identification relevant poles are  located in the region $j\geq 0$).
The pole occurs in the 2nd term of
\eqn{expansion 3} at
\begin{eqnarray}
&& j= \frac{1}{2}(kw+l+\tl)~ \left(\frac{}{} =
  \frac{1}{2}(kw-n+2l) = \frac{1}{2}(kw+n+2\tl)\right)~.
\label{poles} 
\end{eqnarray}
Since only the poles located on the interval between $C_1$ and $C_2$
can contribute, we must impose
\begin{eqnarray}
&& \frac{1}{2}\leq j \leq 
\frac{k+1}{2}\left(= \frac{\kappa-1}{2}\right)~. 
\label{j range}
\end{eqnarray}
Note that this range \eqn{j range} coincides with the one derived in the
bosonic model \cite{HPT} 
(with respect to $\kappa$)
and is strictly smaller than the unitarity
bound \eqn{unitarity j} \cite{DPL} for generic values of $k$. 
This bound also coincides with the one 
required by the analysis of reflection coefficients
(two point functions on sphere) \cite{GK}, and also 
with that obtained from the no-ghost theorem for 
the parent $SL(2;\br)$ theory \cite{MO} (see also \cite{EGP,sl2-old,Pakman}).

We also would like to emphasize that 
the restricted range \eqn{j range} agrees exactly with 
the range of massless matter representations 
$\Chm{(\sNS)}(r,s)$ of ${\cal N}=2$ Liouville theory
discussed in \cite{ES-L}.
In fact under the correspondence of notations
\begin{equation}
j={s\over 2K},\hskip3mm k={N\over K}
\end{equation}
the range $1/2\le j\le (k+1)/2$ maps to
\begin{equation}
K\le s\le N+K
\end{equation}
which is exactly the range of massless representations 
closed under modular transformations.


Recalling the branching relation \eqn{branching D 2}, 
the desired character expansion is obtained as 
\begin{eqnarray}
&& \Zd^{(\sNS)}(\tau)=\frac{C}{k}\sum_{w,n\in\bsz}\,
\sum_{j\in \cJ_{w,n}}\, a(j)
  \chid^{(\sNS)}_{\,\frac{\kappa}{2}-j,m+\frac{k}{2}}(\tau,0)
  \chid^{(\sNS)}_{\,\frac{\kappa}{2}-j,\tm+\frac{k}{2}}(\bar{\tau},0)~,
\label{expansion d} \\
&& \hspace{2cm}
\cJ_{w,n}\equiv \left\lb \frac{1}{2},\,\frac{k+1}{2}\right\rb \cap
\left(\frac{kw-n}{2} + \bz \right)~, 
\label{range j w n} \\
&& \hspace{2cm} a(j) \equiv 
\left\{
\begin{array}{ll}
 1& ~~~ (\frac{1}{2}<j <\frac{k+1}{2}) \\
 \frac{1}{2} & ~~~(j=\frac{1}{2}, \, \frac{k+1}{2})~
\end{array}
\right.
\label{a j}
\end{eqnarray}
where $m$ and $\tm$ are defined as above \eqn{m w n}. 
The factor $a(j)$ is necessary to give a correct weight to the poles 
on the boundary, $j=1/2$ and $(k+1)/2$.\footnote
   {One may object to 
   a fractional factor $1/2$ in the weight of discrete representations in the 
    partition function. However, there exists a character identity
\begin{eqnarray}
\chid^{(\sNS)}_{\,\frac{1}{2},\frac{1}{2}+n}(\tau,z)
 + \chid^{(\sNS)}_{\,\frac{k+1}{2},\frac{k+1}{2}+n}(\tau,z) =
 \chic^{(\sNS)}_{\,j=\frac{1}{2},  m=\frac{1}{2}+n}
(\tau,z) 
\equiv \ch{(\sNS)}{}\left(h=\frac{1}{2k}+\frac{n+n^2}{k},
Q=\frac{2n+1}{k};\tau,z\right)~. \nonumber
\end{eqnarray}
which enables us to write \eqn{expansion d} in a form (set $C=k$) 
\begin{eqnarray}
&& \Zd^{(\sNS)}(\tau)= \sum_{w,n\in\bsz}\,
\sum_{j\in \cJ'_{w,n}}\, 
  \chid^{(\sNS)}_{\,\frac{\kappa}{2}-j,m+\frac{k}{2}}(\tau,0)
  \chid^{(\sNS)}_{\,\frac{\kappa}{2}-j,\tm+\frac{k}{2}}(\bar{\tau},0)
+ (\mbox{terms including continuous rep.})
\nn
&& \hspace{2cm}
\cJ'_{w,n}\equiv \left(  \frac{1}{2},\,\frac{k+1}{2}\right\rb \cap
\left(\frac{kw-n}{2} + \bz \right)~. \nonumber 
\end{eqnarray}
    }
We choose the normalization constant $C=k$.

We finally make a comment with respect to the modular invariance. 
The regularized partition function is written as 
\begin{eqnarray}
&& Z(\tau;\ep) = \frac{1}{2} \sum_{\sigma} \sum_{w,n} 
\left\lb \int_0^{\infty}dp\,
\rho(p,w,n;\ep) \chic^{(\sigma)}_{\,p,m}(\tau,0)
\chic^{(\sigma)}_{\,p,\tm}(\bar{\tau},0) \right. \nn
&& \hspace{2cm} \left. 
+\sum_{j\in \cJ_{w,n}}\, a(j)  
\chid^{(\sigma)}_{\,\frac{\kappa}{2}-j,m+\frac{k}{2}}(\tau,0)
\chid^{(\sigma)}_{\,\frac{\kappa}{2}-j,\tm+\frac{k}{2}}(\bar{\tau},0)
\right\rb~,
\label{reg part fn}
\end{eqnarray}
where we have indicated the dependence on IR cut-off $\ep$ explicitly. 
Strictly speaking this expression is {\em not\/} modular invariant 
because of the non-trivial $p$-dependence 
of the spectral density $\rho(p,w,n;\ep)$ \eqn{evaluation rho},
even though the original formula coming from the path-integral 
\eqn{part fn} appears modular invariant.
In fact, the IR regularization would spoil the modular invariance,
as we often face in general non-compact curved backgrounds. 
In order to recover invariance, the best one can do is to consider the partition function
{\em per unit volume};
\begin{eqnarray}
&& Z(\tau) \equiv \lim_{\ep\,\rightarrow\,+0} \, \frac{Z(\tau;\ep)}
{2\frac{1}{2\pi}\log \ep} = \frac{1}{2}\sum_{\sigma}
\sum_{w,n}\int_0^{\infty}dp\,
\chic^{(\sigma)}_{\,p,m}(\tau,0)
\chic^{(\sigma)}_{\,p,\tm}(\bar{\tau},0)~.
\label{MI part fn} 
\end{eqnarray}
The modular invariance of \eqn{MI part fn} follows from 
that of a free compact boson with a radius $R=\sqrt{2k}$.


To summarize, the partition function is 
decomposed into two parts: 
(1) the continuous part $\Zc(\tau)$, and (2) the discrete
part $\Zd(\tau)$. The continuous part $\Zc(\tau)$
includes dominant contributions proportional to the volume
factor, and correspond to the modes freely propagating in the bulk.
Its precise definition depends on the regularization scheme. 

On the other hand, the discrete part $\Zd(\tau)$ only contains 
representations of (anti-)chiral primaries and their spectral flows
within the range \eqn{j range}.
They describe excitations localized around 
the tip of 2D black-hole that could be identified as the bound states 
\cite{DVV-BH} (see also \cite{RibS}). 
The absence of IR divergence in $\Zd(\tau)$ is in accord with 
this expectation. 
The part of discrete representations is universal:
insensitive to the choice of regularization scheme and stable 
under marginal deformations preserving $\cN=2$ SUSY. 
We will make use of this piece to compute elliptic genus in the next section
which captures the geometrical information of the singular space-time.

It seems that 
a strictly modular invariant partition function is obtained
only after dividing by the infinite volume factor, 
which inevitably contains only the continuous representations. 
All the states appearing in this  partition function 
\eqn{reg part fn} lie above the mass gap $\dsp h\geq 1/(4k)$,
which corresponds to the decoupling of gravity in such a space-time.

~

\newpage
\section{Coupling to RCFT's}

~

In this section we investigate  the type II string vacua of the forms;
\begin{eqnarray}
\br^{d-1,1} \otimes \cM \otimes SL(2;\br)/U(1)~,
\label{type II vacua}
\end{eqnarray}
where $\cM$ is an arbitrary rational $\cN=2$ SCFT with 
$\hat{c}= \hat{c}_{\cM}$.
The criticality condition is 
\begin{eqnarray}
\frac{d}{2} + \hat{c}_{\cM} + \frac{k+2}{k}= 5~,
\label{criticality}
\end{eqnarray}
and we assume $d$ is even. We expect that 
the superconformal system $\cM \otimes SL(2;\br)/U(1)$
describes a non-compact $CY_{\msc{\bf n}}$ with 
$\mbox{\bf n} = 5-d/2$. 
We assume a modular invariant of $\cM$-sector with conformal blocks
$F_I$ as 
\begin{eqnarray}
&& Z_{\cM}(\tau,z)= \frac{1}{2}\sum_{\sigma} Z_{\cM}^{(\sigma)}(\tau,z)
\equiv e^{-2\pi \hat{c}_{\scs{\cM}}\frac{(\sIm\,z)^2}{\tau_2}}
\frac{1}{2}\sum_{\sigma} \sum_{I,\tI}\,N_{I,\tI}
 F_I^{(\sigma)}(\tau,z)  \tilde{F}_{\tI}^{(\sigma)}(\bar{\tau},\bar{z})~, 
\end{eqnarray}
where $z$ is the angle associated to the $U(1)$-charge of 
$\cN=2$ SCA and $\sigma$ denotes the spin structures as before. 
In this paper we use the conventions;
\begin{eqnarray}
&& F_I^{(\stNS)}(\tau,z) \equiv e^{-i\pi Q(I)} 
F_I^{(\sNS)}\left(\tau,z+\frac{1}{2}\right)~,\nn
&& F_I^{(\sR)}(\tau,z) \equiv
 q^{\frac{\hat{c}_{\cM}}{8}}e^{i\pi \hat{c}_{\cM} z}
F_I^{(\sNS)}\left(\tau,z+\frac{\tau}{2}\right)~,\nn
&& F_I^{(\stR)}(\tau,z) \equiv e^{-i\pi Q(I)}
q^{\frac{\hat{c}_{\cM}}{8}}e^{i\pi \hat{c}_{\cM} z}
F_I^{(\sNS)}\left(\tau,z+\frac{\tau}{2}+\frac{1}{2}\right)~,
\label{F os}
\end{eqnarray}
where $Q(I)$ is the $U(1)$-charge of vacuum state of the conformal block
$F_I^{(\sNS)}(\tau,z)$. (Note that the $U(1)$-charge for 
$F_I^{(\sR)}(\tau,z)$ is equal $Q(I)+\hat{c}_{\cM}/2$.)
Because of the rationality of $\cM$ there exists a finite periodicity in
integral spectral flows. We assume  $N_0 \in \bz_{>0}$ to be the minimal
integer such that 
\begin{eqnarray}
 q^{\frac{\hat{c}_{\cM}}{2}m^2} e^{2\pi i \hat{c}_{\cM}m z}
F_I^{(\sNS)}(\tau,z+m\tau+n) = F_I^{(\sNS)}(\tau,z)~,~~~({}^{\forall} I~,~
       {}^{\forall}m, {}^{\forall}n\in N_0\bz)~.
\label{periodicity M}
\end{eqnarray}
Then we have $\hat{c}_{\cM}= M/N_0$ with some positive integer
$M$. Recalling  the criticality condition \eqn{criticality}, we find 
that 
\begin{eqnarray}
k=\frac{N}{K}~,~~~N=N_0 ~\mbox{or}~ 2N_0~,~~~
(\hat{c}_{SL(2;\bsr)/U(1)}=1+\frac{2K}{N}~,~~~ 
1+\frac{2K}{N}+ \frac{M}{N_0}=\mbox{\bf n})~,
\label{N K}
\end{eqnarray}
with some positive integer $K$ \footnote
   {Note that both of the pairs $N_0$, $M$ and $N$, $K$ are {\em not\/}
    necessarily relatively prime. For example, in the case
$$
\br^{3,1} \otimes M_{2n} \otimes SL(2;\br)/U(1)~, ~~~(n\in \bz_{\geq 0})
$$
we find $N_0=N=2n+2$, $M=2n$, $K=n+2$. Therefore, $N_0$, $M$ are not 
relatively prime for any $n$,  and $N$, $K$ are also not for even $n$.
}. Throughout this section we shall assume \eqn{N K} with fixed positive
integers $N$, $K$ in the $SL(2;\br)/U(1)$-sector.

A typical example is the Gepner model \cite{Gepner};
\begin{eqnarray}
\cM = M_{n_1}\otimes \cdots \otimes M_{n_r}~, ~~~
\hat{c}_{\cM} = \sum_{i=1}^r \frac{n_i}{n_i+2}~,
\end{eqnarray}
where $M_n$ is the level $n$ $\cN=2$ minimal model. 
The relevant conformal blocks are the products of minimal characters 
\eqn{minimal character}
\begin{eqnarray}
F^{(\sNS)}_I(\tau,z) = \prod_{i=1}^r \ch{(\sNS)}{\ell_i,m_i}(\tau,z)~,~~~
(I\equiv \left((\ell_1,m_1), \ldots, (\ell_r,m_r)\right))~,
\end{eqnarray}
and clearly we have $N_0= \mbox{L.C.M} \{n_i+2\}$.

We also assume the symmetry under spectral flow of the coefficients of the modular invariant 
$N_{I,\tI}$;
\begin{eqnarray}
 N_{s(I),s(\tI)} = N_{I,\tI} ~, 
\label{N s symmetry}
\end{eqnarray}
where $s\,:\,I\,\rightarrow\, s(I)$ denotes the action of spectral flow  
\begin{eqnarray}
F^{(\sNS)}_{s(I)}(\tau,z) 
= q^{\frac{\hat{c}_{\cM}}{2}} e^{2\pi i \hat{c}_{\cM}z}\,
F^{(\sNS)}_I(\tau,z+\tau)~.
\label{s action F}
\end{eqnarray}

~


\subsection{Modular Invariant Partition Functions per Unit Volume}

~

We first consider the modular invariant partition function 
for the coupled system \\ $\cM \otimes SL(2;\br)/U(1)$ with 
$\hat{c}= \mbox{\bf n}$. 
We assume $N$ and $K$ are relatively prime for the time being. 
Let us recall the partition function of 
the $SL(2;\br)_k/U(1)$-sector 
defined by dividing by the volume factor \eqn{MI part fn}. 
Now in the case of a rational level $k=N/K$, 
the partition function \eqn{MI part fn} can be rewritten in terms of the extended characters
\begin{eqnarray}
&& Z(\tau) = \frac{1}{2} \sum_{\sigma}
\sum_{w_0\in \bsz_{2K}} \sum_{n_0\in \bsz_{N}} \int_0^{\infty}dp\, 
\chic^{(\sigma)}(p, Kn_0-Nw_0  ;  \tau,0)
\chic^{(\sigma)}(p, -Kn_0-Nw_0  ;\bar{\tau},0)~, \nn
&&
\label{MI part fn 2}  
\end{eqnarray}
\begin{eqnarray}
\mathboxit{
\chic^{(\sNS)} (p, m;\tau,z) \equiv \sum_{n\in N \bsz}\,
  \chic^{(\sNS)}_{\,p,{m\over 2K} +n} (\tau,z)  
  \equiv  q^{\frac{Kp^2}{N}} \Th{m}{NK}\left(\tau,\frac{2z}{N}\right)\,
  \frac{\th_3(\tau,z)}{\eta(\tau)^3}
}
\label{chi c}
\end{eqnarray}
This function is identified with the extended massive 
character \eqn{extended massive} introduced in \cite{ES-L};
\begin{eqnarray}
\mathboxit
{
\chic^{(\sNS)} (p, m;\tau,z) 
  = \Ch{(\sNS)}{}\left(h=\frac{Kp^2}{N}+\frac{m^2+K^2}{4NK}, 
   Q=\frac{m}{N}  ;\tau,z \right)
}
\end{eqnarray}
\eqn{MI part fn 2} is derived from the identity 
\begin{eqnarray}
\sum_{w,n\in\bsz}\, 
q^{\frac{1}{4}\left(\sqrt{\frac{K}{N}}n-\sqrt{\frac{N}{K}}w\right)^2}
\bar{q}^{\frac{1}{4}\left(\sqrt{\frac{K}{N}}n+\sqrt{\frac{N}{K}}w\right)^2}
\hskip-1mm=\hskip-2mm \sum_{w_0\in \bsz_{2K}} \sum_{n_0\in\bsz_{N}}\,
\Th{Kn_0-Nw_0}{NK}(\tau,0)\Th{-Kn_0-Nw_0}{NK}(\bar{\tau},0)~, \nn
&&
\end{eqnarray}
and the modular invariant 
has the same form as the level $NK$ theta system 
given in \cite{GQ}. 
$\bz_N$-periodicity under integral spectral flows is easy to see;
\begin{eqnarray}
&&  q^{\frac{\hat{c}}{2}r^2} e^{2\pi i \hat{c} r z}
 \chic^{(\sNS)}(p,m;\tau,z+r\tau+n) = 
 \chic^{(\sNS)} (p, m;\tau,z)~,~{}^{\forall}r,{}^{\forall}n\in N \bz~,~
   (\hat{c}=1+\frac{2K}{N})~. \nn
&&
\label{periodicity chi c}
\end{eqnarray}
$\chic^{(\sigma)}(p,m;\tau,z)$ for other spin structures are defined 
by the 1/2-spectral flows in the same way as \eqn{branching os}.

Now, the task we have to carry out is the chiral projection onto integral
$U(1)$-charges
as in the Gepner models \cite{Gepner} 
while taking account of 
the twisted sectors generated by integral spectral flows (see \cite{EOTY}).  
Because of the periodicities \eqn{periodicity M}, 
\eqn{periodicity chi c}, this is reduced to a $\bz_N$-orbifoldization.
The desired conformal blocks are thus defined as the flow invariant
orbits \cite{EOTY} \footnote
   {In our convention the $\cN=2$ $U(1)$-current in  
    right-mover has been  defined to be 
  $$
     \tJ= -\tpsi^+\tpsi^- -\frac{2}{k} \tJ^3~,
  $$
   and the minus sign of $\bar{z}$ in \eqn{cF cont} reflects  this fact.
    This convention is natural because 
    the winding/KK-momentum $w$, $n$ of the
    compact boson $Y$ are  correctly reinterpreted   as those of 
     the $\cN=2$ $U(1)$-currents $J$, $\tJ$ by using the BRST-invariance. 
     (See \eqn{total current}.)
     } 
\begin{eqnarray}
&&\cF^{(\sNS)}_{I,p,w_0}(\tau,z) = \frac{1}{N}\sum_{a,b\in \bsz_N}\,
q^{\frac{\msc{\bf n}}{2}a^2} e^{2\pi i \msc{\bf n} z a}
\, F_I^{(\sNS)}(\tau,z+a\tau+b) \nn
&&  \hspace{4cm} \times \chic^{(\sNS)} 
  (p, Kn_I-Nw_0;\tau,z+a\tau+b)~, \nn
&&\tilde{\cF}^{(\sNS)}_{\tI,p,w_0}(\bar{\tau},\bar{z}) 
= \frac{1}{N}\sum_{a,b\in \bsz_N}\,
q^{\frac{\msc{\bf n}}{2}a^2} e^{2\pi i \msc{\bf n} \bar{z} a}
\, \tilde{F}_{\tI}^{(\sNS)}(\bar{\tau},\bar{z}+a\bar{\tau}+b) \nn
&&  \hspace{4cm} \times \chic^{(\sNS)} 
  (p, -K n_I-Nw_0;\bar{\tau},-\bar{z}-a\bar{\tau}-b)~,
\label{cF cont} 
\end{eqnarray}
where $n_I  \in \bz_N$ is the solution of the condition
\begin{eqnarray}
&& \frac{Kn_I}{N} + Q(I) \in \bz~, ~~~\frac{K n_I}{N} + Q(\tI) \in \bz ~, ~~~
N_{I,\tI} \neq 0~,
\label{n I}
\end{eqnarray}
which uniquely exist for each $I$, $\tI$ with $N_{I,\tI}\neq 0$
such that $Q(I)-Q(\tI) \in
\bz$, since we have  $\dsp Q(I), Q(\tI) \in \frac{1}{N}\bz $, (${}^{\forall}I,
{}^{\forall}\tI$) and $N$ and $K$ are assumed to be relatively prime.
Especially, the solutions of \eqn{n I} always exist if we assume the 
diagonal modular invariant in the $\cM$-sector. We define 
$\cF^{(\sNS)}_{I,p,w_0}\equiv 0$ if the solution $n_I$ of \eqn{n I} does not 
exist. 
The conformal blocks for other spin structures are defined by 
the $1/2$-spectral flows\footnote
   {We here 
   adopt a somewhat unusual definitions of $\tNS$, $\tR$-conformal blocks
    omitting some phase factors. 
    As an advantage, the supersymmetric conformal blocks \eqn{SUSY conf block} 
    become simpler forms.}
\begin{eqnarray}
&& \cF^{(\stNS)}_{I,p,w_0}(\tau,z)\equiv 
\cF^{(\sNS)}_{I,p,w_0}\left(\tau,z+\frac{1}{2}\right)~,\nn
&& \cF^{(\sR)}_{I,p,w_0}(\tau,z)\equiv 
q^{\frac{\msc{\bf n}}{8}} e^{2\pi i \frac{\msc{\bf n}}{2} z}\,
\cF^{(\sNS)}_{I,p,w_0}\left(\tau,z+\frac{\tau}{2}\right)~, \nn
&&  \cF^{(\stR)}_{I,p,w_0}(\tau,z)\equiv 
q^{\frac{\msc{\bf n}}{8}} e^{2\pi i \frac{\msc{\bf n}}{2} z}\,
\cF^{(\sNS)}_{I,p,w_0}\left(\tau,z+\frac{\tau}{2}+\frac{1}{2}\right)~.
\label{cF cont os}
\end{eqnarray}


Let us next consider the cases when $N$ and $K$ are not relatively
prime. We set 
\begin{eqnarray}
\mbox{G.C.D}\,\{N,\,K\} =\nu~,~~~ N = \bar{N}\nu~,~~~ K=\bar{K}\nu~.
\label{N K nu}
\end{eqnarray}
Then the solutions of the condition \eqn{n I} exist only if 
$\dsp Q(I), Q(\tI) \in \frac{\nu}{N}\bz$, and are not unique:
we must sum over the mod $\bar{N}$ spectral flows $n_I+2 \bar{N}\mu$ 
($\mu\in \bz_{\nu}$).
Namely, \eqn{cF cont} has to  be replaced with 
\begin{eqnarray}
\cF^{(\sNS)}_{I,p,w_0}(\tau,z) &=& \frac{1}{N}\sum_{a,b\in \bsz_N}\,
q^{\frac{\msc{\bf n}}{2}a^2} e^{2\pi i \msc{\bf n} z a}
\, F_I^{(\sNS)}(\tau,z+a\tau+b) \nn
&&  \hspace{2cm} \times \sum_{\mu \in \bsz_{\nu}}\, 
  \chic_{\,(N,K)}^{(\sNS)} 
  (p, K(n_I+2\bar{N}\mu)-Nw_0;\tau,z+a\tau+b) \nn
&=&\frac{1}{N}\sum_{a,b\in \bsz_N}\,
q^{\frac{\msc{\bf n}}{2}a^2} e^{2\pi i \msc{\bf n} z a}
\, F_I^{(\sNS)}(\tau,z+a\tau+b) \, \nn
&&  \hspace{2cm} \times 
  \chic_{\,(\bar{N},\bar{K})}^{(\sNS)} 
  (p, \bar{K}n_I-\bar{N}w_0;\tau,z+a\tau+b)~,
\label{cF cont 2} 
\end{eqnarray}
where we have indicated explicitly the $N$, $K$ dependence 
of the extended characters.


The modular invariant partition function (as the $\sigma$-model on 
$CY_{\msc{\bf n}}$) is obtained as 
\begin{eqnarray}
&& Z(\tau,z) = 
e^{-2\pi \msc{\bf n}\frac{(\sIm\,z)^2}{\tau_2}}
  \frac{1}{2}\sum_{\sigma}\,\sum_{w_0\in \bsz_{2K}}\, 
   \frac{1}{N}\sum_{I,\tI}\, \int_0^{\infty} dp\,
   N_{I,\tI} \cF^{(\sigma)}_{I,p,w_0}(\tau,z)
    \cF^{(\sigma)}_{\tI,p,w_0}(\bar{\tau},\bar{z})~.  
\label{MI part fn sigma}
\end{eqnarray}
We note the invariance (up to phase) under spectral flow of 
the conformal blocks $\cF^{(\sigma)}_{I,p,w_0}$
\begin{eqnarray}
&& q^{\frac{\sn}{2}a^2}e^{2\pi i \sn a z}\,
\cF^{(\sigma)}_{I,p,w_0}(\tau,z+a\tau+b)= \ep_{a,b}(\sigma)
\cF^{(\sigma)}_{I,p,w_0}(\tau,z)~, ~~~({}^{\forall}a,b\in \bz)~, 
\label{cF s symmetry} \\
&& \ep_{a,b}(\NS)=1~,~~\ep_{a,b}(\tNS)=(-1)^{\sn a}~,~~
 \ep_{a,b}(\R)=(-1)^{\sn b}~,~~ \ep_{a,b}(\tR)=(-1)^{\sn (a+b)}~, \nn
\end{eqnarray}
and recall the assumption \eqn{N s symmetry}. 
Then the  factor $1/N$ in \eqn{MI part fn sigma} is necessary to 
remove the $N$-fold overcounting of states.

Incorporating the $\br^{d-1,1}$-sector 
($\dsp \frac{d}{2}+{\bf n}=5$), the supersymmetric conformal blocks are 
constructed as
\begin{eqnarray}
\frac{1}{2} \frac{1}{\tau_2^{\frac{d-2}{4}}\eta(\tau)^{d-2}}\,\sum_{\sigma}\,
  \ep(\sigma)\, 
  \left(\frac{\th_{\lb \sigma \rb}(\tau,z)}{\eta(\tau)}\right)^{\frac{d-2}{2}}
  \, \left(\cF^{(\sigma)}_{I,p,w_0}(\tau,z)+
  \cF^{(\sigma)}_{I,p,w_0}(\tau,-z)\right) ~,
\label{SUSY conf block}
\end{eqnarray}
where $\th_{\lb \sigma \rb}$ again denotes $\th_3$, $\th_4$, $\th_2$,
$i\th_1$ for $\sigma = \NS, \tNS, \R, \tR$ respectively, and we 
set $\ep(\NS)=\ep(\tR)=+1$, $\ep(\tNS)=\ep(\R)=-1$. 
The conformal blocks \eqn{SUSY conf block} actually 
vanish for arbitrary $\tau$, $z$ \cite{HS},
as is consistent with the space-time SUSY.
It is not difficult to confirm that the conformal blocks 
\eqn{SUSY conf block} reproduce the results obtained in \cite{ES1}
for the special cases $\cM=M_{n-2}$.
(Precisely speaking, in the $d=4$ case 
we need some further orbifoldization
in the $SL(2;\br)/U(1)$-sector to reproduce the formula of \cite{ES1}.)

~

\subsection{Elliptic Genera}

~

Let us next study the discrete spectrum of the theory which carries 
geometrical information
of the target space geometry. As we have seen in the previous section,
contributions of continuous representations dominate the partition functions
and it is difficult to isolate the contributions of discrete states of
the theory by inspecting the partition functions.
We thus propose to study the elliptic genera from 
which continuous series decouple and one can clearly see 
the contents of discrete states in the theory.
 
We first recall that the elliptic genera are defined by 
the partition functions in the $\tilde{R}$ sector of the theory
\cite{Witten-E,Witten-E2}, 
\begin{eqnarray}
\cZ(\tau,z) = \tr_{\cH^{(\sR)}_L\otimes \cH^{(\sR)}_R}\, 
(-1)^Fe^{2\pi i z J_0} q^{L_0-\frac{\hat{c}}{8}}
\bar{q}^{\tL_0-\frac{\hat{c}}{8}}~,
\label{elliptic genus 0}
\end{eqnarray}
where $F\equiv F_L-F_R$ denotes the world-sheet fermion number. 
When one sets $z=0$ above, elliptic genus is reduced to the Witten index.
 
It is well-known that, in any rational $\cN=2$ SCFT
the elliptic genus is a good supersymmetric index stable under arbitrary
chiral marginal deformations. Furthermore, it possesses simple modular and spectral flow
properties.  
It is identified as a (weak) Jacobi form \cite{EZ} 
in mathematical terminology
(see, {\em e.g.\/} \cite{Witten-E,KYY}).  

As we shall see in the following, in the case of singular non-compact manifolds
elliptic genera are no longer Jacobi forms and have some complicated modular properties:
they are in general described by Appell functions which feature 
in the study of vector bundles
of higher rank over elliptic curves \cite{Pol,STT}.

The evaluation of elliptic genus is almost parallel 
to the previous analysis of the modular invariant partition 
functions: we just replace the continuous extended characters 
$\chic^{(\sigma)}(p,m;\tau,z)$ 
\eqn{chi c} by the discrete ones $\chid^{(\sigma)}(s,m;\tau,z)$ 
($m\in \bz_{2NK}$) defined by
\begin{equation}
\mathboxit{
\begin{array}{l}
\dsp \chid^{(\sNS)}(s,s+2Kr;\tau,z) \equiv 
\sum_{n\in N\bsz}\,\chid^{(\sNS)}_{\,\frac{s}{2K},\frac{s}{2K}+r+n}(\tau,z)
\equiv  \Chm{(\sNS)}(r,s;\tau,z) ~~~(r\in \bz_N) \\
\dsp \chid^{(\sNS)}(s,m;\tau,z) \equiv 0~,~~~ m\not\equiv s~(\mod\,2K)
\end{array}
}
\label{chi d}
\end{equation}
In this definition
$\chid^{(\sNS)}_{\,j,m}(\tau,z)$ are the branching functions for 
discrete series \eqn{branching D flowed} 
and identified with the $\cN=2$ massless matter characters.  
$\Chm{(\sNS)}(r,s;\tau,z)$ is the massless extended 
character introduced in \cite{ES-L}, given explicitly 
in \eqn{extended massless}.
The extended characters of other spin structures are again defined
by the spectral flows. We note that $\chid^{(\sR)}(s,m)$ 
(and $\chid^{(\stR)}(s,m)$) can take non-zero values only if 
$m\equiv s+K$ ($\mod\, 2K$). 
The discrete part of partition function \eqn{expansion d} 
can be rewritten in terms of the extended characters 
$\chid^{(\sigma)}(s,m;\tau,z)$ in the same way as 
\eqn{MI part fn 2};
\begin{eqnarray}
&& \Zd(\tau) = \frac{1}{2} \sum_{\sigma}
\sum_{w_0\in \bsz_{2K}} \sum_{n_0\in \bsz_{N}} \sum_{s=K}^{N+K}\, a(s) \nn
&& \hspace{1.5cm}\times
\chid^{(\sigma)}(s, Kn_0-Nw_0  ;  \tau,0)
\chid^{(\sigma)}(s, -Kn_0-Nw_0  ;\bar{\tau},0)~,
\label{expansion d 2} \\
&& \hspace{2cm}
a(s)\equiv
\left\{
\begin{array}{ll}
 1& ~~ K+1\leq s \leq N+K-1 \\
 \frac{1}{2} & ~~ s=K,\, N+K~.
\end{array}
\right.
\end{eqnarray}

In the following let us assume 
that $K$ and $N$ are relatively prime for the sake of simplicity.
Performing the $\bz_N$-orbifoldization, we can construct 
the conformal blocks in the same way as \eqn{cF cont}; 
\begin{eqnarray}
&& \cG^{(\sNS)}_{I,s,w_0}(\tau,z) 
= \frac{1}{N}\sum_{a,b\in \bsz_N}\,
q^{\frac{\msc{\bf n}}{2}a^2} e^{2\pi i \msc{\bf n} z a}
\, F_I^{(\sNS)}(\tau,z+a\tau+b) \nn
&&  \hspace{4cm} \times \chid_{\,(N,K)}^{(\sNS)} 
  (s, Kn_I-Nw_0;\tau,z+a\tau+b)~, \nn
&& \tilde{\cG}^{(\sNS)}_{\tI,s,w_0}(\bar{\tau},\bar{z}) 
= \frac{1}{N}\sum_{a,b\in \bsz_N}\,
q^{\frac{\msc{\bf n}}{2}a^2} e^{2\pi i \msc{\bf n} \bar{z} a}
\, \tF_{\tI}^{(\sNS)}(\bar{\tau},\bar{z}+a\bar{\tau}+b) \nn
&&  \hspace{4cm} \times \chid_{\,(N,K)}^{(\sNS)} 
  (s, -Kn_I-Nw_0;\bar{\tau},-\bar{z}-a\bar{\tau}-b)~,
\label{cG}
\end{eqnarray}
where we again assumed \eqn{N K nu}.
The desired $\tR$-parts $\cG^{(\stR)}_{I,s,w_0}(\tau,z)$,
$\tilde{\cG}^{(\stR)}_{\tI,s,w_0}(\bar{\tau},\bar{z})$
are defined by the 1/2-spectral flow as in \eqn{cF cont os}. 
Writing the Witten indices as 
\begin{eqnarray}
\lim_{\bar{z}\,\rightarrow\,0}\, \tilde{\cG}^{(\stR)}_{\tI,s,w_0}
(\bar{\tau},\bar{z})=
\cI_{\tI,s,w_0}~,
\label{cG WI}
\end{eqnarray} 
we obtain the general formula of elliptic genus 
\begin{eqnarray}
&& \cZ(\tau,z)= \frac{1}{N}\sum_{I,\tI}\sum_{s=K}^{N+K}
\sum_{w_0\in \bsz_{2K}}\, a(s) N_{I,\tI} \cI_{\tI,s,w_0}\, 
\cG^{(\stR)}_{I,s,w_0}(\tau,z)~.
\label{elliptic genus}
\end{eqnarray}
For our later calculations it is useful to note the formula of 
Witten index \eqn{Chm WI}, that is,
\begin{eqnarray}
\lim_{z\,\rightarrow\,0}\,\chid_{\,(N,K)}^{(\stR)}
(s,m;\tau,z)
= -\delta^{(2NK)}_{m,s-K}~.
\label{WI chi d}
\end{eqnarray}
It is also useful to introduce
\begin{eqnarray}
\cZ_{N,K}(\tau,z) \equiv -\sum_{s=K}^{N+K}\,
a(s) \chid^{(\stR)}_{\,(N,K)}(s,s-K;\tau,z)~,
\label{cZ N K}
\end{eqnarray}
which describes the $SL(2;\br)/U(1)$-part of elliptic genera 
in the prescription of $\bz_N$-orbifoldization. 
With the help of \eqn{Chm tR} we can further rewrite it as
\begin{eqnarray}
\cZ_{N,K}(\tau,z) &=& -\sum_{s=K}^{N+K}\,
\sum_{n\in \bsz}\,
a(s) \,  \frac{\left(e^{2\pi iz}q^{Nn}\right)^{\frac{s-K}{N}}}
{1-e^{2\pi i z}q^{Nn}}\, 
e^{i \pi  4Kn   z}q^{NK n^2}\, \frac{i\th_1(\tau,z)}{\eta(\tau)^3} \nn
&=& -\sum_{n\in \bsz}\, \frac{e^{i \pi 4Kn z}q^{NKn^2}}
{1-e^{\frac{2\pi i}{N}z} q^n}\, \frac{i\th_1(\tau,z)}{\eta(\tau)^3}
+ \frac{1}{2}\Th{0}{NK}
\left(\tau,\frac{2z}{N}\right)\, \frac{i\th_1(\tau,z)}{\eta(\tau)^3} \nn
&\equiv & -\left\lb\cK_{2NK}\left(\tau,\frac{z}{N},0 
\right) - \frac{1}{2}\Th{0}{NK}
\left(\tau,\frac{2z}{N}\right) \right\rb \,  
\frac{i\th_1(\tau,z)}{\eta(\tau)^3} ~,
\label{cZ N K 2}
\end{eqnarray}
where $\cK_{\ell}(\tau,\nu,\mu)$ is the ``level $\ell$ Appell function''
\cite{Pol,STT} defined by 
\begin{eqnarray}
\cK_{\ell}(\tau,\nu,\mu) \equiv \sum_{m\in \bsz}\, 
\frac{e^{i\pi m^2 \ell \tau +2\pi i m \ell\nu}}
{1-e^{2\pi i (\nu+\mu+m\tau)}}~.
\label{Appell}
\end{eqnarray}

~

We present examples of concrete calculations:

~

\noindent
{\bf Example 1.  Conifold 
($N=K=1$, $\cM$ is trivial,  $\mbox{\bf n}=3$) : }

This is the simplest example and should be identified with the deformed 
conifold (under the T-duality to $\cN=2$ Liouville) \cite{GV}. 
The elliptic genus \eqn{elliptic genus} has a simple form
\begin{eqnarray}
&& \cZ_{\msc{conifold}}(\tau,z) = \cZ_{1,1}(\tau,z)
= -\frac{1}{2}\left(\chid^{(\stR)}_{\,(1,1)}(1,0;\tau,z)
+\chid^{(\stR)}_{\,(1,1)}(2,1;\tau,z)\right)~.
\label{elliptic genus conifold 0}
\end{eqnarray}
We only have the boundary terms of the range \eqn{j range}. 
The following identity is  presented 
in \cite{Odake} and quite useful;
\begin{eqnarray}
\Chm{(\sNS)}(Q=\pm 1;\tau,z) &\equiv  &
 \sum_{n\in \bsz}\,
\frac{q^{n^2-n+1/4} e^{\pm 2\pi i (2n-1) z}}{1+e^{\pm 2\pi i z}q^{n-1/2}} \, 
\frac{\th_3(\tau,z)}{\eta(\tau)^3} \nn
&=& \pm \frac{1}{2}\left(\frac{\Th{1}{3/2}(\tau,2z)}{\eta(\tau)}
-\frac{\Th{-1}{3/2}(\tau,2z)}{\eta(\tau)}\right)
+\frac{1}{2}\Th{1}{1}(\tau,2z) \frac{\th_3(\tau,z)}{\eta(\tau)^3}~, \nn
&&
\label{Odake formula}
\end{eqnarray}
where $\Chm{(\sNS)}(Q;\tau,z)$ is the characters 
of $\hat{c}=3$ extended chiral
algebra \cite{Odake,EOTY} (we use the notation given 
in Appendix C of \cite{ES-L}). 
Applying the $1/2$-spectral flow $\dsp z~\rightarrow~ 
z+\frac{\tau}{2}+\frac{1}{2}$
to both sides of \eqn{Odake formula},
we obtain the identity (see also \cite{STT})
\begin{eqnarray}
&&\chid^{(\stR)}_{\,(1,1)}(1,0;\tau,z) (\equiv
\chid^{(\stR)}_{\,(1,1)}(2,1;\tau,-z)) \equiv \cK_2(\tau,z,0)\, 
\frac{i\th_1(\tau,z)}{\eta(\tau)^3} \nn
&& ~~~ = - \frac{1}{2} \frac{1}{\eta(\tau)}\left(\tTh{-1/2}{3/2}(\tau,2z)
  + \tTh{1/2}{3/2}(\tau,2z)\right) +
\frac{1}{2}\Th{0}{1}(\tau,2z)\, \frac{i\th_1(\tau,z)}{\eta(\tau)^3} \nn
&& ~~~ = -\frac{1}{2}\frac{\th_1(\tau,2z)}{\th_1(\tau,z)}
  + \frac{1}{2}\Th{0}{1}(\tau,2z)\, \frac{i\th_1(\tau,z)}{\eta(\tau)^3}~.
\label{Odake formula 2}
\end{eqnarray}
To derive the last line we used 
the Watson's quintuple product identity \eqn{Watson} 
(see {\em e.g.} \cite{ID}.)
In this way we obtain the simple formula
\begin{eqnarray}
\mathboxit
{
\cZ_{\msc{conifold}}(\tau,z)
= \frac{1}{2} \frac{\th_1(\tau,2z)}{\th_1(\tau,z)}
}
\label{elliptic genus conifold}
\end{eqnarray}

It may be worthwhile to point out the following fact: 
The elliptic genus for the level $n-2$
$\cN=2$ minimal model was calculated in \cite{Witten-E2} by 
using the free field  method in $\cN=2$ Landau-Ginzburg 
theory with superpotential $\dsp W(X)=X^{n}$. 
The result is expressed as 
\begin{eqnarray}
\cZ_{M_{n-2}}(\tau,z)= 
\sum_{\ell=0}^{n-2}\,\ch{(\stR)}{\ell,\ell+1}(\tau,z) 
=- \sum_{\ell=0}^{n-2}\,\ch{(\stR)}{\ell,-(\ell+1)}(\tau,z) 
=\frac{\th_1(\tau, \frac{n-1}{n}z)}
{\th_1(\tau,\frac{1}{n} z)}~,
\label{elliptic genus minimal}
\end{eqnarray}
where $\ch{(\stR)}{\ell,m}(\tau,z)$ is the character of minimal 
model $M_{n-2}$ \eqn{minimal character}.
It is curious to see that the ``analytic continuation'' 
of this formula to the inverse power potential $W(X)=X^{-1}\,(n=-1)$,
which is often used to describe the conformal system of conifold 
\cite{GV}, correctly reproduces our result \eqn{elliptic genus conifold}
(up to normalization).

~

\noindent
{\bf Example 2. ALE($A_{n-1}$) 
($\cM=M_{n-2}$, $N=n$, $K=1$, $\mbox{\bf n}=2$) : }

This is the conformal system first analyzed in \cite{OV} and 
is considered to describe the ALE space obtained 
by deforming the $A_{n-1}$-type 
singularity (in the case of the diagonal modular invariants in
$M_{n-2}$). The formula \eqn{elliptic genus} gives us
\begin{eqnarray}
\mathboxit
{
\cZ_{\msc{ALE}(A_{n-1})}(\tau,z) 
= \sum_{\ell=0}^{n-2}\sum_{\stackrel{m\in \bsz_{2n}}
{\ell+m\equiv 1~(\msc{mod}\, 2)}}
\, \ch{(\stR)}{\ell,m}(\tau,z)\,
\chid_{\,(n,1)}^{(\stR)}(\ell+2,-m;\tau,z)
}
\label{elliptic genus ALE}
\end{eqnarray}
Note that only the representations with 
$2\leq s(\equiv \ell +2) \leq n$ contributes in this case, and  
thus we do not have the boundary terms in contrast with the Example 1.
The Witten index is evaluated as 
\begin{eqnarray}
 \lim_{z\,\rightarrow\,0}\,\cZ_{\msc{ALE}(A_{n-1})}(\tau,z) = n-1~,
\label{WI ALE}
\end{eqnarray}
which reproduces the correct number of non-contractable 2-cycles.
In the special case of Eguchi-Hanson space $\mbox{ALE}(A_1)$ 
the formula \eqn{elliptic genus ALE} is reduced to
\begin{eqnarray}
\cZ_{\msc{ALE}(A_{1})}(\tau,z)& =& 
\chid_{\,(2,1)}^{(\stR)}(2,-1;\tau,z)-
\chid_{\,(2,1)}^{(\stR)}(2,1;\tau,z) \nn
&=& -\ch{\cN=4\,(\stR)}{0}(\ell=0;\tau,z) \nn
&\equiv & -\sum_{n\in\bsz}\, \frac{(-1)^n q^{\frac{1}{2}n(n+1)} 
e^{2\pi i \left(n+\frac{1}{2}\right) z}} {1-e^{2\pi i z}q^n}\,
\frac{i\th_1(\tau,z)}{\eta(\tau)^3}~,  
\end{eqnarray}
where $\ch{\cN=4\,(\stR)}{0}(\ell;\tau,z)$ denotes the level 1 
$\cN=4$ massless  character of spin $\ell(=0,\,1/2)$ (related with 
$\ch{\cN=4\,(\sNS)}{0}(1/2-\ell;\tau,z)$ by the spectral flow) 
\cite{ET}.

Using the formulas \eqn{cZ N K 2} and 
\eqn{elliptic genus minimal}, we can also rewrite
\eqn{elliptic genus ALE} in the form that makes the orbifold structure 
manifest as in \cite{KYY};
\begin{equation}
\mathboxit
{
\begin{array}{lll}
 \cZ_{\msc{ALE}(A_{n-1})}(\tau,z) &=& \dsp
\frac{1}{n}\sum_{a,b\in \bsz_n}\,q^{a^2}e^{i 4\pi a z}\, 
\cZ_{M_{n-2}}(\tau,z+a\tau +b)
\cZ_{n,1}(\tau,z+a\tau +b) \\
&=&  \dsp
-\frac{1}{n}\sum_{a,b\in \bsz_{n}}\, (-1)^{a+b}q^{\frac{a^2}{2}} 
e^{i 2\pi a z}\,
 \frac{\th_1\left(\tau, \frac{n-1}{n}
(z+a\tau+b)\right)}{\th_1\left(\tau,\frac{1}{n}(z+a\tau+b)\right)}\, \\
&& \dsp  \hspace{1cm} \times
\cK_{2n}\left(\tau,\frac{1}{n}(z+a\tau+b), 0\right)
\, \frac{i\th_1(\tau,z)}{\eta(\tau)^3}
\end{array}
}
\label{elliptic genus ALE 2}
\end{equation}
Here we used the relation $\th_1(\tau,z+a\tau+b) = (-1)^{a+b}
q^{-\frac{a^2}{2}} e^{- i 2\pi a z}\, \th_1(\tau,z)$.
Note that the theta function term in \eqn{cZ N K 2}
is eliminated by the $\bz_N$-orbifoldization.

One can also perform similar calculations in the cases of
$CY_4$-singularity of $A_{n-1}$-type ($N=n$, $K=n+1$).
We obtain 
\begin{equation}
\mathboxit{
\begin{array}{lll}
\cZ_{CY_4\,(A_{n-1})}(\tau,z)&=& \dsp  
\sum_{\ell=0}^{n-2}\sum_{\stackrel{m\in \bsz_{2n}}
{\ell+m\equiv 1~(\msc{mod}\, 2)}}
\, \ch{(\stR)}{\ell,m}(\tau,z)\,
\chid_{\,(n,n+1)}^{(\stR)}(\ell+n+2,-m;\tau,z) \\
&=& \dsp 
-\frac{1}{n}\sum_{a,b\in \bsz_{n}}\, (-1)^{a+b}q^{\frac{3a^2}{2}} 
e^{i 6\pi a  z}\,
 \frac{\th_1\left(\tau, \frac{n-1}{n}
(z+a\tau+b)\right)}{\th_1\left(\tau,\frac{1}{n}(z+a\tau+b)\right)}\, \\
&& \dsp \hspace{1cm} \times
\cK_{2n(n+1)}\left(\tau,\frac{1}{n}(z+a\tau+b), 0\right)
\, \frac{i\th_1(\tau,z)}{\eta(\tau)^3}
\end{array}
}
\label{elliptic genus CY4}
\end{equation}

~

\noindent
{\bf Example 3. General cases of non-compact $CY_3$ : }

Lastly we consider the general models of $\mbox{\bf n}=3$, where
$\cM$ is an arbitrary $\cN=2$ RCFT with $\hat{c}_{\cM}<2$.
We make a natural assumption of the ``charge conjugation symmetry'' 
in the $\cM$-sector. Namely, we postulate
\begin{eqnarray}
N_{c(I),c(\tI)}= N_{I,\tI}~,
\label{c c sym}
\end{eqnarray} 
where the charge conjugation $c\,:\, I \,\rightarrow\, c(I)$ 
is defined by 
\begin{eqnarray}
F^{(\sNS)}_{c(I)}(\tau,z)\equiv F^{(\sNS)}_I(\tau,-z)~.
\label{charge conjugation}
\end{eqnarray}
Rather surprisingly we can show that the elliptic genera for these
models have a simple and universal form;
\begin{eqnarray}
\mathboxit{
\cZ(\tau,z)= \frac{\chi}{2}
\left(\frac{\tTh{-1/2}{3/2}(\tau,2z)}{\eta(\tau)}+
\frac{\tTh{1/2}{3/2}(\tau,2z)}{\eta(\tau)}
\right) \equiv \frac{\chi}{2} \frac{\th_1(\tau,2z)}{\th_1(\tau,z)}
}
\label{elliptic genus CY3}
\end{eqnarray}
where $\chi \equiv \lim_{z\,\rightarrow\,0}\, \cZ(\tau,z)$ is the 
Witten index that counts 
the Ramond ground states in the total system.

We sketch how one can derive this formula.
Thanks to the above assumption \eqn{c c sym}, 
the elliptic genus is found to have only contributions 
of the symmetrized forms
$\cG^{(\stR)}_{I,s,w_0}(\tau,z)+\cG^{(\stR)}_{I,s,w_0}(\tau,-z)$. 
Furthermore, because of the $U(1)$-projection 
the $\NS$ conformal blocks $\cG^{(\sNS)}_{I,s,w_0}$ should be 
expanded with positive integer coefficients by the characters of the 
$\hat{c}=3$ extended algebra \cite{Odake}: $\Chm{(\sNS)}(Q=1;\tau,z)$,
$\Chm{(\sNS)}(Q=-1;\tau,z)$, and massive characters
\begin{eqnarray}
&& \Ch{(\sNS)}{}(h,Q=0;\tau,z)\equiv q^{h-1/4} \Th{0}{1}(\tau,2z)
\,\frac{\th_3(\tau,z)}{\eta(\tau)^3}~,~~\nn
&&
\Ch{(\sNS)}{}(h,|Q|=1;\tau,z) \equiv  q^{h-1/2} \Th{1}{1}(\tau,2z)
\,\frac{\th_3(\tau,z)}{\eta(\tau)^3}~.
\end{eqnarray}
(We again use the notations given in Appendix C of \cite{ES-L}.) 
Note that the graviton representation $(h=Q=0)$ cannot occur
owing to the constraint  $K \leq s \leq N+K$ (or
\eqn{j range}). 
Consequently,  recalling the identity \eqn{Odake formula}, \eqn{Odake
formula 2},  one can find 
\begin{eqnarray}
\cG^{(\stR)}_{I,s,w_0}(\tau,z) = \mbox{(integer)} \times 
  \frac{1}{2}\left(\frac{\tTh{-1/2}{3/2}(\tau,2z)}{\eta(\tau)}+
\frac{\tTh{1/2}{3/2}(\tau,2z)}{\eta(\tau)}
\right) + \mbox{(massive part)}~.
\end{eqnarray}
The massive part is generically an infinite sum 
of the terms of the forms 
$q^* \Th{0\,(1)}{1}(\tau,2z)i \th_1(\tau,z)/\eta(\tau)^3$ and hence 
odd functions of $z$. It does not contribute because of the above remark.
In this way we arrive at  the general formula \eqn{elliptic genus CY3}.

It has been shown in \cite{KYY} that 
the elliptic genera for arbitrary Gepner models (or the LG-orbifolds)
of compact $CY_3$ can be written in  the form
\begin{eqnarray}
\cZ_{\msc{Gepner}}(\tau,z) = (h_{1,2}-h_{1,1})  
\left(\frac{\tTh{-1/2}{3/2}(\tau,2z)}{\eta(\tau)}+
\frac{\tTh{1/2}{3/2}(\tau,2z)}{\eta(\tau)}
\right)~,
\label{elliptic genus CY3 2}
\end{eqnarray} 
where $h_{1,2}$, $h_{1,1}$ are the numbers of 
$(c,c)$, $(c,a)$-type chiral primaries of $h=\tilde{h}=1/2$
respectively, which should 
be identified with the Hodge numbers of $CY_3$.
Our result \eqn{elliptic genus CY3} is the generalization of this
formula to the non-compact models. Note that 
$\chi = 2(h_{1,2}-h_{1,1})$ is an even number for any 
compact $CY_3$, while $\chi$ in \eqn{elliptic genus CY3} is allowed to 
be odd. Recall the conifold case,  Example 1.

~

A few remarks are in order:

~

\noindent
{\bf 1.}
The elliptic genus includes contributions of Ramond ground states 
that are naively supposed to describe massless excitations in string 
theory. 
However, to identify  them with the massless spectrum we must 
take  account of the GSO condition for spin structures.
The simplest way to do so is to look for the corresponding NS 
(anti-)chiral states with $\dsp h=\frac{1}{2}|Q|=\frac{1}{2}$ . 
For example, in the $\mbox{ALE}(A_{n-1})$ case (Example 2), 
we have $n-1$ Ramond ground states and all of them correspond 
to massless string states. They are identified as  each of 
normalizable deformations of $A_{n-1}$-singularity (``moduli''). 
However, even though the $A_{n-1}$-model 
of the Calabi-Yau 4-fold $\mbox{\bf n}=4$ ($N=n$, $K=n+1$)
still has the Witten index $\cZ(\tau,0)=n-1$, they cannot 
define $\NS$ massless states at all.  This feature has its origin in 
the restriction $K\leq s \leq N+K$. 
We thus find the inequality
\begin{eqnarray}
 h \equiv \frac{s}{2N} \geq \frac{K}{2N} \equiv \frac{1}{2}+\frac{1}{2n}
  > \frac{1}{2}~,
\end{eqnarray}
for the chiral primary states in the $SL(2;\br)/U(1)$-sector, implying 
no massless states appear in the superstring spectrum. 
These missing massless  states would  be associated to 
the {\em non-normalizable\/} deformations of $A_{n-1}$-singularity
(``coupling constants'').  In the Calabi-Yau 3-fold $\mbox{\bf n}=3$ case the aspect of 
massless states is more complex: the half of them appears 
as massless states in the closed string spectrum and the remaining ones 
are missing. These aspects of marginal fields in 
singular $CY_{\msc{\bf n}}$ have been discussed in \cite{GVW,ShV,GKP,Pelc}.

~

\noindent
{\bf 2. }
As we already mentioned, the elliptic genera of $\cN=2$ RCFT's for 
compact $CY_{\msc{\bf n}}$ are known to be the (weak) Jacobi form with 
weight 0 and index $\mbox{\bf n}/2$. This means that 
the elliptic genera have the following properties;
\begin{eqnarray}
&&q^{\frac{\msc{\bf n}}{2} a^2} e^{2\pi i \msc{\bf n} a z}\,
\cZ(\tau, z+a\tau+b) =  (-1)^{\sn (a+b)}\,
\cZ(\tau,z) ~,  ~~~ ({}^{\forall}a,b\in \bz)~,
\label{quasi-periodicity} \\
&&\cZ(\tau+1,z) = \cZ(\tau,z) ~, 
\label{cZ T} \\
&& \cZ\left(-\frac{1}{\tau}, \frac{z}{\tau}\right) 
= e^{i\pi \msc{\bf n}\frac{z^2}{\tau}} \cZ(\tau,z)~.
\label{cZ S}
\end{eqnarray}
It is easy to confirm that our elliptic genera for non-compact models 
\eqn{elliptic genus} satisfy \eqn{quasi-periodicity} and 
\eqn{cZ T}. However, the third condition for the S-transformation
is not necessarily obeyed. 
For the $\mbox{\bf n}=3$ cases, \eqn{cZ S} is also 
satisfied because of the general formula \eqn{elliptic genus CY3}. 
However, in the $\mbox{\bf n}=2,4$ cases, the
elliptic genera are proportional to the Appell function which 
transforms as \cite{STT}
\begin{eqnarray}
&&{\cal K}_{\ell}(-{1\over \tau},{\nu\over \tau},{\mu\over \tau})
=\tau e^{i\pi\ell{\nu^2-\mu^2\over\tau}}
{\cal K}_{\ell}(\tau,\nu,\mu)\nonumber \\
&&\hskip3cm +\tau\sum_{a=0}^{\ell-1}e^{i\pi{\ell\over \tau}
(\nu+{a\over \ell}\tau)^2}
\Phi(\ell\tau,\ell\mu-a\tau)\theta_3(\ell\tau,\ell\nu+a\tau)
\end{eqnarray}
where
\begin{equation}
\Phi(\tau,\mu)=-{i\over 2\sqrt{-i\tau}}-{1\over 2}
\int^{\infty}_{-\infty} dx\, e^{-\pi x^2}
{\sinh(\pi x\sqrt{-i\tau}(1+2{\mu\over \tau}))\over\sinh(\pi x\sqrt{-i\tau})}
\end{equation}
The above transformation law corresponds 
to the mixing of discrete and continuous 
representations
in ${\cal N}=2$ Liouville theory \cite{ET,Odake,Miki,ES-L} (see Appendix B).
Appell functions \eqn{Appell} 
are interpreted as sections of higher rank vector bundles 
over elliptic curves as opposed to theta-functions which are sections of
line bundles \cite{Pol}.
It is interesting to see if one can  
achieve a precise geometrical 
interpretation of the elliptic genera in the non-compact space-time.


~

\section{Summary}

~

In this paper we have confirmed  the correspondence between 
$SL(2;\br)_k/U(1)$ supercoset and ${\cal N}=2$ Liouville 
theory and computed the elliptic genera for various
singular space-times. We summarize the main ingredients of this paper. 
\begin{enumerate}
 \item Partition functions of $SL(2;\br)_k/U(1)$ theories
are decomposed into two pieces: 
(1) the part consisting of continuous (massive) representations and 
(2) the part consisting of discrete (massless matter) representations.
 \item The continuous part is proportional to the volume
       factor, since it describes the propagating modes in the bulk,
       and gives the leading contribution to the partition function.
       It seems that strictly modular invariant partition functions 
       are obtained 
       after the division by the infinite volume factor. 
       Then they contain only the continuous representations (massive modes),
       and reproduce the results obtained previously in \cite{ES1}. 
 \item The discrete part describes excitations localized around the tip 
       of cigar \cite{DVV-BH} and thus appears without the 
       volume factor. Embedded in superstring vacua
       it could be interpreted as contributions from 
       massless matter fields corresponding to the deformations of 
       Calabi-Yau singularities. 
 \item Continuous representations do not contribute to the elliptic genera
       and thus elliptic genus clearly exhibits the contributions of the discrete states.
       In generic cases (of $\hat{c}_{\msc{tot}}=2,4$) the elliptic 
       genera possess complex modular behaviors, and they 
       are not Jacobi forms (section of line bundles)
       but sections of higher rank vector 
       bundles. On the other hand, in models with 
        $\hat{c}_{\msc{tot}}=3$ the elliptic genera behave in the same 
        way as rational conformal theories. 
 \item When embedded in superstring vacua by means of the Gepner-like method,  
       the extended characters defined in \cite{ES-L} emerges   
       quite naturally in continuous and discrete
       series (see \eqn{chi c}, \eqn{chi d}).
       We have also confirmed that discrete representations in 
       ${\cal N}=2$ Liouville theory closed under modular 
       transformations are mapped to unitary
       discrete representations in the range $1/2\le j\le (k+1)/2$ 
       which appear in the regularized partition function of
       $SL(2,{\bf R})_k/U(1)$ theory. This justifies our Ansatz 
       for the basis of Ishibashi states of ${\cal N}=2$ Liouville
       theory we have proposed in \cite{ES-L}.
 \item  It appears quite likely that 
       ${\cal N}=2$ Liouville and $SL(2,{\bf R})_k/U(1)$ theories are
       in fact exactly 
       mapped into each other (T-duality) and will 
       essentially be one and the same theory with identical 
       physical contents. 
       This is gratifying since in the Liouville approach 
        it has been extremely   
       difficult to incorporate the effects of the Liouville potential terms 
       non-perturbatively into the theory.  
       We note that in the $SL(2,{\bf R})_k/U(1)$ supercoset theory, 
       on the other hand, the space-time is  
       curved into a 2D black hole but the (cosmological constant) parameter 
       $\mu$ does not appear explicitly. Thus it seems that 
        $SL(2,{\bf R})_k/U(1)$ theory 
       has deformed the space-time by absorbing the Liouville potential terms. 
       Agreement of our Liouville results \cite{ES-L} with those of  
       $SL(2,{\bf R})_k/U(1)$ theory is encouraging and 
       indicates that we have incorporated properly the effects of 
       Liouville potential terms in the analysis.

\end{enumerate}

~


\section*{Acknowledgements}
\indent

We are grateful to A. Taormina for her correspondence about 
the higher level Appell functions. We also would like to thank 
Y. Nakayama for valuable discussions. 

The research of T. E. and Y. S. is partially  supported by 
Japanese Ministry of Education, 
Culture, Sports, Science and Technology.

\newpage
\section*{Appendix A ~ Notations and Some Useful Formulas}
\setcounter{equation}{0}
\def\theequation{A.\arabic{equation}}

~

\noindent
{\bf 1. Theta functions}

We here summarize our notations of theta functions.
We set  $q\equiv e^{2\pi i \tau}$ and  $y\equiv e^{2\pi i z}$, 
 \begin{eqnarray}
&& \theta_1(\tau,z) =i\sum_{n=-\infty}^{\infty}(-1)^n q^{(n-1/2)^2/2} y^{n-1/2}
  \equiv 2 \sin(\pi z)q^{1/8}\prod_{m=1}^{\infty}
    (1-q^m)(1-yq^m)(1-y^{-1}q^m), \nn
&&  \theta_2(\tau,z)=\sum_{n=-\infty}^{\infty} q^{(n-1/2)^2/2} y^{n-1/2}
  \equiv 2 \cos(\pi z)q^{1/8}\prod_{m=1}^{\infty}
    (1-q^m)(1+yq^m)(1+y^{-1}q^m)~, \nn
&& \theta_3(\tau,z)=\sum_{n=-\infty}^{\infty} q^{n^2/2} y^{n}
  \equiv \prod_{m=1}^{\infty}
    (1-q^m)(1+yq^{m-1/2})(1+y^{-1}q^{m-1/2})~, \nn
&& \theta_4(\tau,z)=\sum_{n=-\infty}^{\infty}(-1)^n q^{n^2/2} y^{n}
  \equiv \prod_{m=1}^{\infty}
    (1-q^m)(1-yq^{m-1/2})(1-y^{-1}q^{m-1/2}) ~,
\end{eqnarray}
 \begin{eqnarray}
 &&\Th{m}{k}(\tau,z)=\sum_{n=-\infty}^{\infty}
 q^{k(n+\frac{m}{2k})^2}y^{k(n+\frac{m}{2k})} ~, \nn
 && \tTh{m}{k}(\tau,z)=\sum_{n=-\infty}^{\infty} (-1)^n
 q^{k(n+\frac{m}{2k})^2}y^{k(n+\frac{m}{2k})}~. \nn   
&&\eta(\tau)=q^{1/24}\prod_{n=1}^{\infty}(1-q^n)~.
 \end{eqnarray}

~

\noindent
{\bf 2. Character Formulas for the $\cN=2$ Minimal Models}

The easiest way to represent the character formulas of 
the level $k$ $\cN=2$ minimal model $(\hat{c}=k/(k+2))$ is to use its
realization as the coset
$\dsp \frac{SU(2)_k\times SO(2)_1}{U(1)_{k+2}}$. We then have 
the following branching relation;
\begin{eqnarray}
&& \chi_{\ell}^{(k)}(\tau,w)\Th{s}{2}(\tau,w-z)
=\sum_{\stackrel{m\in \bsz_{2(k+2)}}{\ell+m+s\in 2\bsz}} \chi_m^{\ell,s}
(\tau,z)\Th{m}{k+2}(\tau,w-2z/(k+2))~, \nn
&& \chi^{\ell,s}_m(\tau,z) \equiv  0~, ~~~ \mbox{for $\ell+m+s \in 2\bz+1$}~,
\label{branching minimal}
\end{eqnarray}
where $\chi_{\ell}^{(k)}(\tau,z)$ is the spin $\ell/2$ character of 
$SU(2)_k$;
\begin{eqnarray}
&&\chi^{(k)}_{\ell}(\tau, z) 
=\frac{\Th{\ell+1}{k+2}(\tau,z)-\Th{-\ell-1}{k+2}(\tau,z)}
                        {\Th{1}{2}(\tau,z)-\Th{-1}{2}(\tau,z)}
\equiv \sum_{m \in \bsz_{2k}}\, c^{(k)}_{\ell,m}(\tau)\Th{m}{k}(\tau,z)~.
\label{SU(2) character}
\end{eqnarray}
The branching function $\chi^{\ell,s}_m(\tau,z)$ 
is explicitly calculated as follows;
\begin{equation}
\chi_m^{\ell,s}(\tau,z)=\sum_{r\in \bsz_k}c^{(k)}_{\ell, m-s+4r}(\tau)
\Th{2m+(k+2)(-s+4r)}{2k(k+2)}(\tau,z/(k+2))~.
\end{equation}
Then, the desired character formulas are written as 
\begin{eqnarray}
&& \ch{(\sNS)}{\ell,m}(\tau,z) = \chi^{\ell,0}_m(\tau,z)
+\chi^{\ell,2}_m(\tau,z)~, \nn
&& \ch{(\stNS)}{\ell,m}(\tau,z) = \chi^{\ell,0}_m(\tau,z)
-\chi^{\ell,2}_m(\tau,z)\equiv 
e^{-i\pi\frac{m}{k+2}}\ch{(\sNS)}{\ell,m}\left(\tau,z+\frac{1}{2}\right)~, \nn
&& \ch{(\sR)}{\ell,m}(\tau,z) = \chi^{\ell,1}_m(\tau,z)
+\chi^{\ell,3}_m(\tau,z) \equiv 
q^{\frac{k}{8(k+2)}}y^{\frac{k}{2(k+2)}}
\ch{(\sNS)}{\ell,m+1}\left(\tau,z+\frac{\tau}{2}\right)~, \nn
&& \ch{(\stR)}{\ell,m}(\tau,z) = \chi^{\ell,1}_m(\tau,z)
-\chi^{\ell,3}_m(\tau,z) \equiv
- e^{-i\pi\frac{m+1}{k+2}}q^{\frac{k}{8(k+2)}}y^{\frac{k}{2(k+2)}}
\ch{(\sNS)}{\ell,m+1}\left(\tau,z+\frac{1}{2}+\frac{\tau}{2}\right)~. \nn
&& \label{minimal character}
\end{eqnarray}
By definition, we restrict to $\ell+m \in 2\bz$ in $\NS$ and
$\tNS$ sectors, and to $\ell+m \in 2\bz+1$ in $\R$ and $\tR$
sectors.

~

\noindent
{\bf 3. Useful identity}

\begin{eqnarray}
&& \prod_{n=1}^{\infty}(1-q^n)(1-yq^n)(1-y^{-1}q^{n-1})
(1-y^2q^{2n-1})(1-y^{-2}q^{2n-1}) \nonumber \\
&& \hspace{2cm}
= \sum_{m\in\bsz} \left(y^{3m}-y^{-3m-1}\right) q^{\frac{1}{2}m(3m+1)},
~~~\nonumber \\
&& \hspace{2cm}
\Longleftrightarrow ~
\frac{1}{\eta(\tau)}\left(\tTh{1/2}{3/2}(\tau,2z)+
\tTh{-1/2}{3/2}(\tau,2z)\right) =
\frac{\th_1(\tau,2z)}{\th_1(\tau,z)} \nonumber \\
&& \hspace{4cm}
(\mbox{Watson's quintuple product identity})~. \label{Watson} 
\end{eqnarray}

The following identity is often useful in checking the modular invariance 
\begin{equation}
\frac{\left(\Im\,\left(\frac{u}{\tau}\right)\right)^2}
{\Im\,\left(-\frac{1}{\tau}\right)}
= \frac{(\Im \,u)^2}{\Im\, \tau} + 
i\frac{u^2}{2\tau}-i\frac{\bar{u}^2}{2\bar{\tau}}.
\label{anomaly factor}
\end{equation}
We also note that the combination $|u|^2/\tau_2$ is  modular invariant.

~

\section*{Appendix B ~ $\cN=2$ (Extended) Character Formulas for $\hat{c}>1$}
\setcounter{equation}{0}
\def\theequation{B.\arabic{equation}}

~

We denote the conformal weight and $U(1)$-charge of the highest weight state
as $h$, $Q$ and again set $q\equiv e^{2\pi i\tau}$, $y\equiv e^{2\pi i z}$.
The irreducible characters of $\cN=2$ SCA with $\hat{c}>1$ are 
summarized as follows \cite{Dobrev};
\begin{description}
 \item[(1) massive representations] :
\begin{eqnarray}
\ch{(\sNS)}{}(h,Q;\tau,z) = q^{h-(\hat{c}-1)/8}\,y^Q\,
\frac{\theta_3(\tau,z)}{\eta(\tau)^3}~, ~~~(h>|Q|/2,~~0\leq |Q| 
< \hat{c}-1)~.
\label{massive character}
\end{eqnarray}
\item[(2) massless matter representations] : 
\begin{eqnarray}
\chm{(\sNS)}(Q;\tau,z)=q^{\frac{|Q|}{2}-(\hat{c}-1)/8}y^Q\,
\frac{1}{1+y^{\msc{sgn}(Q)}q^{1/2}}
\, \frac{\theta_3(\tau,z)}{\eta(\tau)^3}~.
\label{massless character 1}
\end{eqnarray}
They correspond to the (anti-)chiral primary state with $h=|Q|/2$, ~
($0<|Q|<\hat{c}$).
\item[(3) graviton representation] : 
\begin{eqnarray}
\chg{(\sNS)}(\tau,z)=q^{-(\hat{c}-1)/8}\,
\frac{1-q}{(1+y q^{1/2})(1+y^{-1}q^{1/2})} \,
\frac{\theta_3(\tau,z)}{\eta(\tau)^3}~.
\label{massless character 2}
\end{eqnarray}
They correspond to the vacuum $h=Q=0$, 
which is the unique state being both chiral and anti-chiral primary. 
\end{description}

More general unitary representations are generated by the integral 
spectral flows and classified in \cite{BFK}. 
The spectral flow generator $U_{\eta}$
with a real parameter $\eta$ is defined by 
\begin{eqnarray}
&& U_{\eta}^{-1}L_m U_{\eta}
=L_m+\eta J_m+\frac{\hat{c}}{2}\eta^2\delta_{m,0}~,\nn
&& U_{\eta}^{-1}J_m U_{\eta} = J_m+ \hat{c}\eta \delta_{m,0}~,\nn
&& U_{\eta}^{-1}G^{\pm}_r U_{\eta}
= G^{\pm}_{r\pm \eta}~.
\label{spectral flow}
\end{eqnarray}
Half-integral spectral flows $\dsp \eta \in \frac{1}{2}+\bz$ 
intertwine the NS and R sector characters,
while the integral spectral flows $\eta=n\in \bz$
keep the spin structure. 
The spectrally flowed characters  are given by 
\begin{eqnarray}
\ch{(\sNS)}{*}(*,n;\tau,z) \equiv 
q^{\frac{\hat{c}}{2}n^2}
y^{\hat{c}n}\, \ch{(\sNS)}{*}(*;\tau,z+n\tau)~, ~~~(n\in \bz)~,
\label{flowed character}
\end{eqnarray}
where $\ch{(\sNS)}{*}(*;\tau,z)$ is the abbreviated notation of
\eqn{massive character}-\eqn{massless character 2}.

For the theory of $\hat{c}=1+2K/N$ ($N,K\in \bz_{>0}$), 
we introduce the ``extended characters'' \cite{ES-L} 
which should be the characters of unitary representations 
of the extended chiral algebra defined by adding the spectral flow
generators $U_{\pm N}$ \cite{ET,Odake,EOTY,HS}. 
\begin{eqnarray}
\Ch{(\sNS)}{}(h,\al;\tau,z)
&=&
\sum_{n\in r+N\bsz}\, q^{\frac{\hat{c}}{2}n^2}y^{\hat{c}n}\,
\ch{(\sNS)}{}\left(h_0,Q=\frac{\al_0}{N};\tau,z+n\tau\right) \nn
&\equiv & q^{p^2/2}\Th{\al}{NK}\left(\tau,\frac{2z}{N}\right)\,
\frac{\th_3(\tau,z)}{\eta(\tau)^3}~,  
\label{extended massive}\\
&& (h\equiv h_0 + \frac{r\al_0}{N}+\frac{Kr^2}{N} \equiv 
\frac{p^2}{2}+\frac{\al^2+K^2}{4NK}~,~~~ 
\al \equiv \al_0 + 2Kr)~, \nn
\Chm{(\sNS)}(r,s;\tau,z)
&=& 
\sum_{n\in r+N\bsz}\, q^{\frac{\hat{c}}{2}n^2}y^{\hat{c}n}\,
\chm{(\sNS)}\left(Q=\frac{s}{N};\tau,z+n\tau\right)~, \nn
&\equiv &\sum_{m\in \bsz}\, \frac{\left(yq^{N\left(m+\frac{2r+1}{2N}\right)}
\right)^{\frac{s-K}{N}}}{1+yq^{N\left(m+\frac{2r+1}{2N}\right)}}\,
y^{2K\left(m+\frac{2r+1}{2N}\right)} q^{NK\left(m+\frac{2r+1}{2N}\right)^2}
\,\frac{\theta_3(\tau,z)}{\eta(\tau)^3}~,
\label{extended massless} \\
\Chg{(\sNS)}(r;\tau,z)
&= & 
\sum_{n\in r+N\bsz}\, q^{\frac{\hat{c}}{2}n^2}y^{\hat{c}n}\,
\chg{(\sNS)}\left(\tau,z+n\tau\right)~, \nn
&\equiv &q^{-\frac{K}{4N}}\,\sum_{m\in\bsz}\, 
q^{NK\left(m+\frac{r}{N}\right)^2+N\left(m+\frac{2r-1}{2N}\right)} 
y^{2K\left(m+\frac{r}{N}\right)+1}\, \nn
&& \hspace{1cm} \times
\frac{1-q}{\left(1+yq^{N\left(m+\frac{2r+1}{2N}\right)}\right)
\left(1+yq^{N\left(m+\frac{2r-1}{2N}\right)}\right)}
\, \frac{\theta_3(\tau,z)}{\eta(\tau)^3}~,
\label{extended graviton}
\end{eqnarray}
where the ranges of parameters $r$, $\al$, $s$ are given as
\begin{eqnarray}
&& r\in \bz_{N}~,~~~
\al \in \bz_{2NK}~, ~~~
1 \le s \le N+2K-1~, ~~~(s\in \bz)~.
\label{range r s}
\end{eqnarray}
In the calculations of elliptic genera, 
we use the following formula ($\dsp r\in \frac{1}{2}+\bz_N$)
\begin{eqnarray}
 \Chm{(\stR)}(r,s;\tau,z) &\equiv &e^{-i\pi \frac{s+K(2r-1)}{N}}
q^{\frac{\hat{c}}{8}}y^{\frac{\hat{c}}{2}}\,
\Chm{(\sNS)}\left(r-\frac{1}{2},s;\tau,z+\frac{\tau}{2}+\frac{1}{2}\right)
\nn
&=& \sum_{m\in \bsz}\,\frac{\left(yq^{N\left(m+\frac{2r+1}{2N}\right)}
\right)^{\frac{s-K}{N}}}{1-yq^{N\left(m+\frac{2r+1}{2N}\right)}}\,
y^{2K\left(m+\frac{2r+1}{2N}\right)} q^{NK\left(m+\frac{2r+1}{2N}\right)^2}\,
\frac{i\th_1(\tau,z)}{\eta(\tau)^3}~,
\label{Chm tR}
\end{eqnarray}
which yields the Witten index
\begin{eqnarray}
 \lim_{z\,\rightarrow\,0}\,\Chm{(\stR)}(r,s;\tau,z) = -\delta^{(N)}_{r, -1/2}~.
\label{Chm WI}
\end{eqnarray}

For convenience of readers we also present the modular transformation 
formulas given in \cite{ES-L} in terms of 
the new notations \eqn{chi c}, \eqn{chi d}.
They are written as 
\begin{eqnarray}
&&\chic^{(\sNS)}\left(p,m;-\frac{1}{\tau}, \frac{z}{\tau}\right)
=e^{i\pi \hat{c}\frac{z^2}{\tau}}
\frac{2}{N} \sum_{m'\in \bsz_{2NK}}\, e^{-2\pi i \frac{m m'}{2NK}}\, \nn
&& \hspace{2cm} \times
\int_0^{\infty}dp'\, \cos\left(2\pi \frac{2K}{N}pp'\right)\,
\chic^{(\sNS)}(p',m';\tau,z)~, 
\label{S cont}
\end{eqnarray}
\begin{eqnarray}
&&\chid^{(\sNS)}\left(s,m;-\frac{1}{\tau},\frac{z}{\tau}\right)
=e^{i\pi \hat{c}\frac{z^2}{\tau}}
\, \left\lb 
\frac{1}{N}\sum_{m'\in \bsz_{2NK}}\, e^{-2\pi i \frac{m m'}{2NK}}
\right.  \nn
&& \hspace{1cm} \times \int_0^{\infty} dp'\,
\frac{\cosh\left(2\pi \frac{N-(s-K)}{N}p'\right)
+ e^{2\pi i\frac{m'}{2K}}
\cosh\left(2\pi \frac{s-K}{N}p'\right)}
{2\left|
\cosh \pi \left(p'+i\frac{m'}{2K}\right)
\right|^2} \, \chic^{(\sNS)}(p',m';\tau,z) \nn
&& \hspace{1cm} 
+ \frac{i}{N} \sum_{s'=K+1}^{N+K-1}\,\sum_{m'\in \bsz_{2NK}}\, 
e^{2\pi i \frac{(s-K)(s'-K)-m m'}{2NK}}\,  
\chid^{(\sNS)}(s',m';\tau,z) \nn
&& \hspace{1cm} \left. 
+ \frac{i}{2N} \sum_{m'\in \bsz_{2NK}}\, e^{-2\pi i \frac{m m'}{2NK}}\,
\left\{
\chid^{(\sNS)}(K,m';\tau,z) - \chid^{(\sNS)}(N+K,m'+N;\tau,z)
\right\}
\right\rb~. \nn
&&
\label{S discrete} 
\end{eqnarray}


~


\section*{Appendix C ~ Explicit Calculation of Partition Function}
\setcounter{equation}{0}
\def\theequation{C.\arabic{equation}}

~

In this Appendix we present the explicit derivation of 
\eqn{part fn} by the path-integration. 
This is almost parallel to the analysis of the bosonic models 
given in \cite{GawK,Gaw}. 
We define the world-sheet torus by the identifications  
$(w,\bar{w}) \sim (w+2\pi,\bar{w}+2\pi)
\sim(w+2\pi \tau, \bar{w}+2\pi \bar{\tau})$ ($\tau\equiv
\tau_1+i\tau_2$, $\tau_2>0$, 
and use the convention $z=e^{iw}$, $\bar{z}=e^{-i\bar{w}}$).
We call the cycles defined by these two identifications
as the $\al$ and $\beta$-cycles as usual.

The desired partition function is written as 
\begin{eqnarray}
Z(\tau) = \int \cD\lb g, A, \psi^{\pm}, \tpsi^{\pm}\rb\,
e^{-\kappa S_{\msc{gWZW}}(g,A) - S_{\psi}(\psi^{\pm},\tpsi^{\pm}, A)}~.
\label{part fn 0}
\end{eqnarray}
Although the Euclidean 2D BH is manifestly positive definite
(since the time-like $U(1)$ is gauged away), 
the calculation of partition function could be subtle 
due to the Lorentzian signature in the parent $SL(2;\br)$ theory. 
Therefore, it is better to start with the Wick rotated model 
$H^+_3/\br$, where $H^+_3 \cong SL(2;\bc)/SU(2)$ is the Euclidean
$AdS_3$. 
The Wick rotation is defined by the replacement; 
$
g\in SL(2;\br) \, \rightarrow \, g\in SL(2;\bc)/SU(2)
$
and the gauge field 
$ \dsp
A \equiv \left(A_{\bar{z}}d\bar{z}+ A_z dz\right)\frac{\sigma_2}{2}
$
should be  regarded as a hermitian 1-form.

To calculate the partition function \eqn{part fn 0} 
it is convenient to reexpress the gauged WZW action \eqn{gWZW action} in 
the form
\begin{eqnarray}
&& S_{\msc{gWZW}} (g,A) = S_{\msc{WZW}}(h_Lg h_R) 
-S_{\msc{WZW}}(h_L h_R^{-1})  \equiv 
S^{(A)}_{\msc{gWZW}} (g,h_L,h_R)~,
\label{gWZW action 2}
\\
&& A_{\bar{z}} \frac{\sigma_2}{2}= \partial_{\bar{z}} h_L h_L^{-1}~,~~~
A_{z} \frac{\sigma_2}{2}= \partial_{z} h_R h_R^{-1}~, ~~~
\end{eqnarray}
where we used the abbreviated notation $S_{\msc{WZW}}(g)\equiv 
S_{\msc{WZW}}^{SL(2;\bsr)}(g)$. After the Wick rotation we 
must suppose  $h_L=h_R^{\dag}(\equiv h) \in \exp\left(\bc \sigma_2\right)$.
We also introduce the vector-like gauged WZW action
\begin{eqnarray}
&& S^{(V)}_{\msc{gWZW}} (g,h_L,h_R) \equiv S_{\msc{WZW}}(h_Lgh_R) 
- S_{\msc{WZW}}(h_Lh_R)~. 
\label{vector action} 
\end{eqnarray}
We can parameterize $h(\equiv h_L \equiv h_R^{\dag})$ as 
\begin{eqnarray}
h= e^{(X+iY)\frac{\sigma_2}{2}} h^u ~, ~~ h^u \equiv 
e^{i\Phi^u\frac{\sigma_2}{2}}~,
\label{parameterization h}
\end{eqnarray}
where $\Phi^u(w,\bar{w})$ is associated with the modulus of holomorphic
line bundle; $u\equiv s_1\tau-s_2 \in \mbox{Jac}(\Sigma) \cong \Sigma$, 
$(0\leq s_1,s_2 <1)$, conventionally defined as 
\begin{eqnarray}
\Phi^u(w,\bar{w})= \frac{i}{2\tau_2}\left\{
(w\bar{\tau}-\bar{w}\tau)s_1+(\bar{w}-w)s_2\right\}~,~
\label{Phi u}
\end{eqnarray} 
It is a real harmonic function 
satisfying the twisted boundary conditions
\begin{eqnarray}
\Phi^u(w+2\pi,\bar{w}+2\pi)=\Phi^u(w,\bar{w}) + 2\pi s_1~,~~~
\Phi^u(w+2\pi\tau,\bar{w}+2\pi\bar{\tau})=\Phi^u(w,\bar{w}) + 2\pi s_2~.
\end{eqnarray}
Real scalar fields $X$, $Y$ correspond 
to the axial ($\br_A$) and vector ($U(1)_V$) 
gauge transformations respectively.
Using the Polyakov-Wiegmann identity 
\begin{eqnarray}
&& S^{(A)}_{\msc{gWZW}} (\Om g\Om^{\dag},\Om^{-1}h,\Om^{\dag\,-1}h^{\dag})  
= S^{(V)}_{\msc{gWZW}}(g,h,h^{\dag}) 
- S^{(A)}_{\msc{gWZW}}(\Om^{-1}\Om^{\dag}, h, h^{\dag\,-1})~,
\label{PW}
\end{eqnarray} 
and the gauge invariance of path-integral measure 
$\cD (\Om g \Om^{\dag})=\cD g$,
we can rewrite the partition function \eqn{part fn 0} as follows
(after dividing by the gauge volume $\dsp \int \cD X$);
\begin{eqnarray}
Z(\tau) &=& \int_{\Sigma} d^2u \,\int \cD\lb g, Y, \psi^{\pm},\tpsi^{\pm},
  b,\tilde{b}, c, \tilde{c}\rb\,
 e^{-\kappa S^{(V)}(g,h^u,h^{u\,\dag}) + (\kappa-2) 
S^{(A)}(e^{iY\sigma_2}, h^u,h^{u\,\dag\,-1}) } \nn 
&& \hspace{2cm} \times e^{-S_{\psi}(\psi^{\pm},\tpsi^{\pm}, a^u) -
S_{\msc{gh}}(b,\tilde{b},c,\tilde{c})} \nn
&\equiv & \int_{\Sigma} d^2u \, Z_g(\tau,u) Z_Y(\tau,u) Z_{\psi}(\tau,u)
Z_{\msc{gh}}(\tau)~,
\label{part fn 1} \\
&& (a_{\bar{w}}^u \equiv i\partial_{\bar{w}}\Phi^u(w,\bar{w}) \equiv 
    \frac{u}{2\tau_2}~,~~~ 
a_{w}^u \equiv -i\partial_{w}\bar{\Phi}^{u}(w,\bar{w}) \equiv 
    \frac{\bar{u}}{2\tau_2})~. \nonumber
\end{eqnarray}
Here $b$, $c$ ($\tilde{b}$, $\tilde{c}$) are the spin (1,0) ghost 
system to rewrite the Jacobian of path integral measure. 
The level shift $\kappa \,\rightarrow\,\kappa-2 (\equiv k)$ for 
the action $S^{(A)}(e^{iY\sigma_2}, h^u,h^{u\,\dag\,-1})$
in \eqn{part fn 1} is owing to the chiral anomaly of the fermion
determinant, regularized so that it is anomaly free along the axial 
direction.

We next evaluate each sector separately:
\begin{itemize}
\item
{\bf $H_3^+$-sector : } 

This non-trivial sector has been already evaluated 
in \cite{GawK,Gaw} (see the comment below);
\begin{eqnarray}
Z_g(\tau,u) \equiv \int \cD g\, e^{-\kappa S^{(V)}(g, h^u,h^{u\,\dag})} 
\propto \frac{e^{2\pi \frac{(\sIm\,u)^2}{\tau_2}}}{\sqrt{\tau_2}
|\th_1(\tau,u)|^2}~.
\label{Z g}
\end{eqnarray}
$Z_g(\tau,u)$ is indeed modular invariant, 
especially under the S-transformation 
$\tau\,\rightarrow\,-1/\tau$, $u\,\rightarrow\,u/\tau$. 
Note that the ``anomaly factor'' $e^{2\pi \frac{(\sIm\,u)^2}{\tau_2}}$ 
remedies the S-invariance thanks to the identity \eqn{anomaly factor} 
as mentioned in Appendix B of \cite{MO}.

\item
{\bf $U(1)_V$-sector : }

$Y$ is the coordinate along $U(1)_V$-direction
($i\br \sigma_2$) and thus compact; $Y\sim Y+2\pi$. 
The relevant world-sheet action is calculated as
\begin{eqnarray}
S_Y(Y;u)&\equiv& 
-kS_{\msc{WZW}}(h^u e^{iY\sigma_2} h^{u\,\dag\,-1}) 
+kS_{\msc{WZW}}(h^uh^{u\,\dag}) \nn
&=& \frac{k}{\pi}\int d^2w\, 
\left|\partial_{\bar{w}}Y-ia^u_{\bar{w}}\right|^2 = \frac{k}{\pi}\int d^2w\, 
\left|\partial_{\bar{w}}Y^u\right|^2~.
\end{eqnarray} 
In the last line we set $Y^u\equiv Y +\Phi^u$, 
which satisfies the twisted boundary conditions;
\begin{eqnarray}
&&Y^u(w+2\pi,\bar{w}+2\pi)=Y^u(w,\bar{w})+ 2\pi (m+s_1) ~, ~~~(m\in \bz) \nn
&&Y^u(w+2\pi\tau,\bar{w}+2\pi\bar{\tau})=Y^u(w,\bar{w})+ 2\pi (n+s_2)
~, ~~~(n\in \bz)~.
\end{eqnarray}
Rescaling  
the twisted boson $Y^u$ as $Y^u\, \rightarrow\, Y^u/\sqrt{2k}$,
we arrive at the theory of a twisted compact boson
with radius $R=\sqrt{2k}$. Therefore, the relevant partition
function becomes
\begin{eqnarray}
Z_Y(\tau,u)&=& \int \cD Y^u\, e^{-\frac{1}{2\pi} \int d^2w\, 
\left|\partial_w Y^u\right|^2} \nn
&\propto&
\frac{1}{\sqrt{\tau_2}\left|\eta(\tau)\right|^2}\,
\sum_{m,n\in \bsz}\, \exp\left(-\frac{\pi k}{\tau_2}
\left|(m+s_1)\tau-(n+s_2)\right|^2\right)~.
\label{Z Y} 
\end{eqnarray}
Again this is manifestly modular invariant.

\item
{\bf fermion and ghost sectors :}

The remaining fermionic sectors are easy to evaluate.
They are the standard fermion determinants with anti-periodic and
periodic boundary conditions respectively (for the NS sector of $\psi^{\pm}$) 
\begin{eqnarray}
Z^{(\sNS)}_{\psi}(\tau,u) &=& \int \cD\lb\psi^{\pm},\tpsi^{\pm}\rb\,
e^{-S_{\psi}(\psi^{\pm},\tpsi^{\pm},a^u)} =
  e^{-2\pi \frac{(\sIm\,u)^2}{\tau_2}}
\frac{\left|\th_3(\tau,u)\right|^2}{\left|\eta(\tau)\right|^2}~,
\label{Z psi} \\
Z_{\msc{gh}}(\tau) &=& \int \cD\lb b, \tilde{b}, c, \tilde{c}\rb\,
e^{-S_{\msc{gh}}(b,\tilde{b},c,\tilde{c})} = \tau_2\left|\eta(\tau)\right|^4~.
\label{Z gh}
\end{eqnarray}
The factor $e^{-2\pi \frac{(\sIm\,u)^2}{\tau_2}}$ included 
in \eqn{Z psi} is the correct anomaly factor to assure the modular invariance. 
\end{itemize}

~

Gathering all the contributions \eqn{Z g}, \eqn{Z Y}, 
\eqn{Z psi} and \eqn{Z gh}, we finally obtain the 
desired partition function \eqn{part fn}.

~

We further make a comment on the several path-integral formulas for
the $H_3^+$ (gauged) WZW models presented in \cite{GawK,Gaw};
\begin{eqnarray}
&& \int \cD g\, e^{-\kappa S_{\msc{WZW}}(h^ugh^{u\,\dag})} 
\equiv  \tr\, \left(q^{L_0-\frac{c_g}{24}}
\bar{q}^{\tilde{L}_0-\frac{c_g}{24}}\,
 e^{2\pi i (u j^3_0 -\bar{u} \tilde{j}^3_0)} \right) 
\propto \frac{e^{-(\kappa -2)\pi \frac{(\sIm\,u)^2}{\tau_2}}}{\sqrt{\tau_2}
|\th_1(\tau,u)|^2}~, \label{Gaw formula 1} \\
&& \int \cD g\, e^{-\kappa S^{(V)}(g,h^u,h^{u\,\dag})} 
\propto \frac{e^{2\pi \frac{(\sIm\,u)^2}{\tau_2}}}{\sqrt{\tau_2}
|\th_1(\tau,u)|^2}~, \label{Gaw formula 2} \\ 
&& \int \cD g\, e^{-\kappa S^{(A)}(g,h^u,h^{u\,\dag})} 
\propto \frac{e^{2\pi \frac{(\sIm\,u)^2}{\tau_2} 
- \pi \kappa \frac{|u|^2}{\tau_2}}}{\sqrt{\tau_2}
|\th_1(\tau,u)|^2}~, \label{Gaw formula 3} 
\end{eqnarray}
where 
$h^u$ is defined in \eqn{parameterization h}.
The second and third formulas \eqn{Gaw formula 2}, \eqn{Gaw formula 3}
are modular invariant, but the first one \eqn{Gaw formula 1} is not,
although it has a natural interpretation as the trace in 
the operator calculus.

The third formula \eqn{Gaw formula 3} is the easiest to prove. 
The axial action $S^{(A)}(g,h^u,h^{u\,\dag})$ can be rewritten as 
a complete quadratic form by taking suitable coordinates on $H_3^+$ 
\cite{Gaw}. The relevant calculation is then reduced to 
successive Gaussian integrals and the chiral anomaly formulas 
($\phi$ is a non-compact real scalar along the $\br \sigma_2$-direction 
and $v$ is a complex  scalar);
\begin{eqnarray}
&&\int \cD g \, e^{-\kappa S^{(A)}(g,h^u,h^{u\,\dag})} =
\int \cD\lb \phi, v, \bar{v}\rb\, 
e^{-\frac{\kappa}{\pi} \int d^2w\, \left\{
(\partial_w \phi + a_w^u)(\partial_{\bar{w}} \phi + a_{\bar{w}}^u)
+(\partial_w+\partial_w \phi + a_w^u)\bar{v}
(\partial_{\bar{w}}+\partial_{\bar{w}} \phi + a_{\bar{w}}^u)v
\right\}} \nn
&&= \int \cD \phi \, e^{-\frac{\kappa}{\pi} \int d^2w\, 
\left|\partial_w \phi + a^u_w\right|^2}
\times \det \left((\partial_{\bar{w}}+\partial_{\bar{w}}\phi
+a^u_{\bar{w}})^{\dag}
(\partial_{\bar{w}}+\partial_{\bar{w}}\phi+a^u_{\bar{w}})
\right)^{-1}  \nn
&&= \int \cD \phi \, e^{-\frac{\kappa}{\pi} \int d^2w\, 
\left|\partial_w \phi + a^u_w\right|^2 
+\frac{2}{\pi}\int d^2w \,\left|\partial_w \phi\right|^2
+\frac{1}{2\pi i} \int \phi R}
\times \det \left((\partial_{\bar{w}}+a^u_{\bar{w}})^{\dag}
(\partial_{\bar{w}}+a^u_{\bar{w}})\right)^{-1} \nn
&&\propto  \frac{e^{-\pi \kappa \frac{|u|^2}{\tau_2}}}
{\sqrt{\tau_2}|\eta(\tau)|^2}
\times \left(\frac{e^{-2\pi \frac{(\sIm \, u)^2}{\tau_2}}|\th_1(\tau,u)|^2}
 {|\eta(\tau)|^2}\right)^{-1} =
\frac{1}{\sqrt{\tau_2}} \frac{e^{2\pi \frac{(\sIm \, u)^2}{\tau_2} 
-\pi \kappa\frac{|u|^2}{\tau_2}}}
{|\th_1(\tau,u)|^2}~.
\label{proof of Gaw formula 3} 
\end{eqnarray}
We have thus obtained the formula \eqn{Gaw formula 3}.
In the last line the path-integration of $\phi$ is evaluated as
\begin{eqnarray}
&&\int \cD \phi\, e^{-\frac{\kappa-2}{\pi} \int d^2w\, 
\partial_w \phi\partial_{\bar{w}}\phi - \frac{\kappa}{\pi} \int d^2w\,
\left(\partial_w \phi a^u_{\bar{w}}+ \partial_{\bar{w}}\phi a^u_{w}\right)
-\frac{\kappa}{\pi} \int d^2w\, \left|a_w^u\right|^2} \nn
&& ~~~ = e^{-\pi \kappa \frac{|u|^2}{\tau_2}}
\int \cD \phi\, e^{-\frac{\kappa-2}{\pi} \int d^2w\, 
\partial_w \phi\partial_{\bar{w}}\phi + \frac{i\kappa}{2\pi} \int 
\tilde{a}^u\wedge \, d\phi} \propto 
\frac{e^{-\pi \kappa \frac{|u|^2}{\tau_2}}}
{\sqrt{\tau_2}|\eta(\tau)|^2}~,
\end{eqnarray}
where we set $\tilde{a}^u = a^u_{\bar{w}}d\bar{w}- a^u_w dw
\equiv i d\Phi^u$, 
which satisfies 
\begin{eqnarray}
\oint_{\al} \tilde{a}^u = 2\pi i s_1~, ~~~ 
\oint_{\beta} \tilde{a}^u = 2\pi i s_2~.
\end{eqnarray}
Since $\phi$ is non-compact, we have $\dsp \oint_{\al} d\phi 
= \oint_{\beta} d\phi =0$, and hence the linear term of $\phi$ 
does not contribute. 

The remaining formulas \eqn{Gaw formula 1} and \eqn{Gaw formula 2}
are readily derived from \eqn{Gaw formula 3} by using the properties  
$\dsp S_{\msc{WZW}}(h^uh^{u\,\dag})= \frac{\pi (\Im \, u)^2}{\tau_2}$, 
$\dsp S_{\msc{WZW}}(h^uh^{u\,\dag\,-1})= -\frac{\pi (\Re \, u)^2}{\tau_2}$.

~



\newpage
\begin{thebibliography}{99}






\bibitem{GV}
D.~Ghoshal and C.~Vafa,
Nucl.\ Phys.\ B {\bf 453}, 121 (1995)
[arXiv:hep-th/9506122].


\bibitem{OV}
H.~Ooguri and C.~Vafa,
Nucl.\ Phys.\ B {\bf 463}, 55 (1996)
[arXiv:hep-th/9511164].

\bibitem{ABKS}
O.~Aharony, M.~Berkooz, D.~Kutasov and N.~Seiberg,
JHEP {\bf 9810}, 004 (1998)
[arXiv:hep-th/9808149].


\bibitem{GKP}
A.~Giveon, D.~Kutasov and O.~Pelc,
JHEP {\bf 9910}, 035 (1999)
[arXiv:hep-th/9907178].


\bibitem{GK}
A.~Giveon and D.~Kutasov,
JHEP {\bf 9910}, 034 (1999)
[arXiv:hep-th/9909110];
A.~Giveon and D.~Kutasov,
JHEP {\bf 0001}, 023 (2000)
[arXiv:hep-th/9911039].


\bibitem{Pelc}
O.~Pelc,
JHEP {\bf 0003}, 012 (2000)
[arXiv:hep-th/0001054].


\bibitem{ES1}
T.~Eguchi and Y.~Sugawara,
Nucl.\ Phys.\ B {\bf 577}, 3 (2000)
[arXiv:hep-th/0002100].


\bibitem{Mizoguchi}
S.~Mizoguchi,
JHEP {\bf 0004}, 014 (2000)
[arXiv:hep-th/0003053].


\bibitem{Mizoguchi2}
S.~Mizoguchi,
arXiv:hep-th/0009240.


\bibitem{Yamaguchi}
S.~Yamaguchi,
Nucl.\ Phys.\ B {\bf 594}, 190 (2001)
[arXiv:hep-th/0007069];
Phys.\ Lett.\ B {\bf 509}, 346 (2001)
[arXiv:hep-th/0102176];
JHEP {\bf 0201}, 023 (2002)
[arXiv:hep-th/0112004].


\bibitem{NN}
M.~Naka and M.~Nozaki,
Nucl.\ Phys.\ B {\bf 599}, 334 (2001)
[arXiv:hep-th/0010002].


\bibitem{HK2}
K.~Hori and A.~Kapustin,
JHEP {\bf 0211}, 038 (2002)
[arXiv:hep-th/0203147].


\bibitem{Murthy}
S.~Murthy,
JHEP {\bf 0311}, 056 (2003)
[arXiv:hep-th/0305197].


\bibitem{GS2}
M.~Gutperle and A.~Strominger,
Phys.\ Rev.\ D {\bf 67}, 126002 (2003)
[arXiv:hep-th/0301038].


\bibitem{ST}
A.~Strominger and T.~Takayanagi,
Adv.\ Theor.\ Math.\ Phys.\  {\bf 7}, 2 (2003)
[arXiv:hep-th/0303221].


\bibitem{Schomerus}
V.~Schomerus,
arXiv:hep-th/0306026.


\bibitem{Nakayama}
Y.~Nakayama,
arXiv:hep-th/0402009.


\bibitem{ES-L}
T.~Eguchi and Y.~Sugawara,
JHEP {\bf 0401}, 025 (2004)
[arXiv:hep-th/0311141].


\bibitem{FZZ}
V.~Fateev, A.~B.~Zamolodchikov and A.~B.~Zamolodchikov,
arXiv:hep-th/0001012.


\bibitem{Teschner}
J.~Teschner,
arXiv:hep-th/0009138.




\bibitem{ZZ}
A.~B.~Zamolodchikov and A.~B.~Zamolodchikov,
arXiv:hep-th/0101152.



\bibitem{ASY}
C.~Ahn, M.~Stanishkov and M.~Yamamoto,
arXiv:hep-th/0311169.


\bibitem{KS}
Y.~Kazama and H.~Suzuki,
Nucl.\ Phys.\ B {\bf 321}, 232 (1989).



\bibitem{FZZ2}
V.~Fateev, A.~B.~Zamolodchikov and A.~B.~Zamolodchikov,
unpublished.

\bibitem{HK1}
K.~Hori and A.~Kapustin,
JHEP {\bf 0108}, 045 (2001)
[arXiv:hep-th/0104202].


\bibitem{Tong}
D.~Tong,
JHEP {\bf 0304}, 031 (2003)
[arXiv:hep-th/0303151].



\bibitem{HPT}
A.~Hanany, N.~Prezas and J.~Troost,
JHEP {\bf 0204}, 014 (2002)
[arXiv:hep-th/0202129].

\bibitem{MOS}
J.~M.~Maldacena, H.~Ooguri and J.~Son,
J.\ Math.\ Phys.\  {\bf 42}, 2961 (2001)
[arXiv:hep-th/0005183].

\bibitem{IKP}
D.~Israel, C.~Kounnas and M.~P.~Petropoulos,
JHEP {\bf 0310}, 028 (2003)
[arXiv:hep-th/0306053].



\bibitem{GawK}
K.~Gawedzki and A.~Kupiainen,
Nucl.\ Phys.\ B {\bf 320}, 625 (1989).

\bibitem{Gaw}
K.~Gawedzki,
arXiv:hep-th/9110076.


\bibitem{2DBH}
E. Witten, 
Phys. Rev. {\bf D44} (1991) 314;
G. Mandal, A. Sengupta and S. Wadia, 
Mod. Phys. Lett. {\bf A6} (1991) 1685;
I. Bars and D. Nemeschansky, 
Nucl. Phys. {\bf B348} (1991) 89;
S. Elizur, A. Forge and E. Rabinovici, 
Nucl. Phys. {\bf B359} (1991) 581.







\bibitem{DVV-BH}
R.~Dijkgraaf, H.~Verlinde and E.~Verlinde,
Nucl.\ Phys.\ B {\bf 371}, 269 (1992).




\bibitem{Witten-E}
E.~Witten,
Commun.\ Math.\ Phys.\  {\bf 109}, 525 (1987).


\bibitem{Pol}
A. ~Polishchuk,
arXiv:math.AG/9810084.

\bibitem{STT}
A.~M.~Semikhatov, A.~Taormina and I.~Y.~Tipunin,
arXiv:math.qa/0311314.




\bibitem{IPT}
D.~Israel, A.~Pakman and J.~Troost,
arXiv:hep-th/0402085.


\bibitem{KarS}
D.~Karabali and H.~J.~Schnitzer,
Nucl.\ Phys.\ B {\bf 329}, 649 (1990).


\bibitem{DPL}
L.~J.~Dixon, M.~E.~Peskin and J.~Lykken,
Nucl.\ Phys.\ B {\bf 325}, 329 (1989).


\bibitem{FST}
B.~L.~Feigin, A.~M.~Semikhatov and I.~Y.~Tipunin,
J.\ Math.\ Phys.\  {\bf 39}, 3865 (1998)
[arXiv:hep-th/9701043];
B.~L.~Feigin, A.~M.~Semikhatov, V.~A.~Sirota and I.~Y.~Tipunin,
Nucl.\ Phys.\ B {\bf 536}, 617 (1998)
[arXiv:hep-th/9805179].



\bibitem{BFK}
W.~Boucher, D.~Friedan and A.~Kent,
Phys.\ Lett.\ B {\bf 172}, 316 (1986).




\bibitem{MO}
J.~M.~Maldacena and H.~Ooguri,
J.\ Math.\ Phys.\  {\bf 42} (2001) 2929, hep-th/0001053.


\bibitem{EGP}
J.~M.~Evans, M.~R.~Gaberdiel and M.~J.~Perry,
Nucl.\ Phys.\ B {\bf 535}, 152 (1998)
[arXiv:hep-th/9806024].

\bibitem{sl2-old}
J. Balog, L. O'Raifeartaigh, P. Forgacs  and A. Wipf,
Nucl. Phys. {\bf B325} (1989) 225;
P. Petropoulos, 
Phys. Lett. {\bf B236} (1990) 151;
N. Mohammedi,
Int. J. of Mod. Phys. {\bf A5} (1990) 3201;
I. Bars and D. Nemeschansky, 
Nucl. Phys. {\bf B348}, 89 (1991);
S. Hwang, 
Nucl. Phys. {\bf B354} (1991) 100.


\bibitem{Pakman}
A.~Pakman,
JHEP {\bf 0301} (2003) 077
[arXiv:hep-th/0301110].




\bibitem{RibS}
S.~Ribault and V.~Schomerus,
arXiv:hep-th/0310024.


\bibitem{Gepner}
D.~Gepner,
Phys.\ Lett.\ B {\bf 199}, 380 (1987);
Nucl.\ Phys.\ B {\bf 296}, 757 (1988).


\bibitem{GQ}
D.~Gepner and Z.~A.~Qiu,
Nucl.\ Phys.\ B {\bf 285}, 423 (1987).



\bibitem{EOTY}
T.~Eguchi, H.~Ooguri, A.~Taormina and S.~K.~Yang,
Nucl.\ Phys.\ B {\bf 315}, 193 (1989).



\bibitem{HS}
Y.~Hikida and Y.~Sugawara,
JHEP {\bf 0210}, 067 (2002)
[arXiv:hep-th/0207124].



\bibitem{Witten-E2}
E.~Witten,
Int.\ J.\ Mod.\ Phys.\ A {\bf 9}, 4783 (1994)
[arXiv:hep-th/9304026].



\bibitem{EZ}
M. Eichler and D. Zagier, 
{\it``The Theory of Jacobi Forms,''}
(Birkh\"{a}user, 1985).


\bibitem{KYY}
T.~Kawai, Y.~Yamada and S.~K.~Yang,
Nucl.\ Phys.\ B {\bf 414}, 191 (1994)
[arXiv:hep-th/9306096].



\bibitem{Odake}
S.~Odake,
Mod.\ Phys.\ Lett.\ A {\bf 4}, 557 (1989);
Int.\ J.\ Mod.\ Phys.\ A {\bf 5}, 897 (1990).


\bibitem{ID}
C. Itzykson, J.-M. Drouffe, 
{\it ``Statistical Field Theory,'' }
(Cambridge), {\bf vol 2}, Appendix 9. A.  

\bibitem{ET}
T.~Eguchi and A.~Taormina,
Phys.\ Lett.\ B {\bf 200}, 315 (1988);
Phys.\ Lett.\ B {\bf 210}, 125 (1988).



\bibitem{GVW}
S.~Gukov, C.~Vafa and E.~Witten,
Nucl.\ Phys.\ B {\bf 584}, 69 (2000)
[Erratum-ibid.\ B {\bf 608}, 477 (2001)]
[arXiv:hep-th/9906070].


\bibitem{ShV}
A.~D.~Shapere and C.~Vafa,
arXiv:hep-th/9910182.


\bibitem{Miki}
K.~Miki,
Int.\ J.\ Mod.\ Phys.\ A {\bf 5}, 1293 (1990).



\bibitem{Dobrev}
V.~K.~Dobrev,
Phys.\ Lett.\ B {\bf 186}, 43 (1987);
E.~Kiritsis,
Int.\ J.\ Mod.\ Phys.\ A {\bf 3}, 1871 (1988).
























\end{thebibliography}
\end{document}